\documentclass[12pt,amsmath]{mn2e}
\usepackage{graphicx}   
\usepackage{times}
\usepackage{float} 
\usepackage{lscape}  
\usepackage{aastex_hack}
\usepackage[export]{adjustbox}
\usepackage[authoryear]{natbib}   
\usepackage{amsmath} 
\usepackage{setspace}
\usepackage{layout} 
\usepackage{wrapfig}
\usepackage{subfigure}
\usepackage{tabularx}
\usepackage{lscape}
\usepackage{rotating}
\usepackage{amssymb}
\usepackage{multirow}
\usepackage[usenames,dvipsnames,svgnames,table]{xcolor}
\pdfoutput=1

\begin{document}

\title[Faraday Depth Structure of NGC\,612]{Revealing the Faraday Depth Structure of Radio Galaxy NGC\,612 with Broad-Band Radio Polarimetric Observations}
\author[J. F. Kaczmarek et al.]
{J. F.~Kaczmarek$^{1,2,3}$\thanks{email: jane.kaczmarek@csiro.au}
C. R.~Purcell,$^{2,3}$ B. M.~Gaensler,$^{2,4}$ X.~Sun,$^{2,5}$ 
\newauthor  S. P.~O'Sullivan,$^6$ N.\,M.~McClure-Griffiths$^7$  \\
	$^1$CSIRO Astronomy and Space Science, Australia Telescope National Facility, Box 76, Epping, NSW 1710, Australia \\ 
    $^2$Sydney Institute for Astronomy, School of Physics, The University of Sydney, NSW 2006 Australia \\
 	$^3$Research Centre for Astronomy, Astrophysics, and Astrophotonics, Macquarie University, NSW 2109, Australia  \\
 	$^4$Dunlap Institute for Astronomy and Astrophysics, University of Toronto, 50 St. George Street, Toronto, ON M5S 3H4, Canada \\
 	$^5$Department of Astronomy, Yunnan University, and Key Laboratory of Astroparticle Physics of Yunnan Province, Kunming, 650091, China\\
 	$^6$Hamburger Sternwarte, Universit\"{a}t Hamburg, Gojenbergsweg 112, Hamburg 21029, Germany\\
 	$^7$Research School of Astronomy and Astrophysics, Australian National University, Canberra, ACT 2611, Australia 
}
\maketitle
\begin{abstract} 
We present full-polarisation, broadband observations of the radio galaxy NGC\,612 (PKS B0131-637) from 1.3 to 3.1\,GHz using the Australia Telescope Compact Array. The relatively large angular scale of the radio galaxy makes it a good candidate with which to investigate the polarisation mechanisms responsible for the observed Faraday depth structure. By fitting complex polarisation models to the polarised spectrum of each pixel, we find that a single polarisation component can adequately describe the observed signal for the majority of the radio galaxy. While we cannot definitively rule out internal Faraday rotation, we argue that the bulk of the Faraday rotation is taking place in a thin skin that girts the polarised emission. Using minimum energy estimates, we find an implied total magnetic field strength of $4.2\,\mu$G. 

\end{abstract} 

\section{Introduction}
The synchrotron emission associated with radio galaxy lobes can be used to dissect the evolutionary history of the host galaxy as well as that of the surrounding intergalactic medium. Jets that are launched from a central supermassive black hole create, and subsequently inflate, radio lobes \citep{Begelman1984, Xu2010}. As the lobes expand, they have the potential to interact with the surrounding environment, possibly implanting signatures from the interaction into the generated synchrotron plasma. Additionally, these interactions can enrich the surrounding medium with large amounts of energy and metals \citep{McNamara2009, Aguirre2001, Reuland2007}. However, the relative amount of thermal material present in radio lobes is poorly constrained, and it has been shown that the lobes are predominantly inflated with non-thermal, synchrotron-emitting plasma \citep{Begelman1984}. Evidence for thermal material distributed throughout the volumes of radio lobes has been argued in the case of some of the most well-studied radio galaxies, Centaurus A \citep{OSullivan2012} and Fornax A \citep{Fomalont1989}. The key to detecting thermal material in radio lobes may lie in the detailed analysis of the Faraday depth structure, as the Faraday depth encodes the thermal electron density in addition to the line-of-sight~magnetic field strength. Therefore, by studying the polarised emission associated with radio lobes we can gain insight towards the origin and density of thermal gas in radio galaxies and the surrounding intergalactic medium.
  
Differentiating between polarisation contributions that are internal and those that are external to the source has proven difficult. The vast majority of investigations into the nature of large-scale polarisation signatures associated with radio galaxies have found distributions that appear nonhomogeneous (e.g. \citealt{Bonafede2010, Govoni2010}). The irregular distributions of Faraday depth structures have been largely attributed to the superposition of intervening material along the line of sight. \citet{Laing1988}, \citet{Kronberg2008} and \citet{Guidetti2010} successfully modelled rotation measures associated with radio galaxies as a result of foreground emission from the large scale, diffuse intracluster medium (ICM). 
 
In contrast, other authors have argued that a significant portion of the observed RM is intrinsic to the radio lobe itself. In the latter case, there is even more debate as to where in the lobe the Faraday rotation takes place. \citet{RudnickBlundell2003}, \citet{Guidetti2011} and \citet{Guidetti2012} make the case that the Faraday rotating material forms a thin skin encompassing the purely synchrotron lobes, though later work by \citet{Ensslin2003} demonstrated limitations in the null experiment performed by \citet{RudnickBlundell2003}.  In contrast, \citet{OSullivan2013} successfully fit their observations by modelling radio lobes as a mixture of relativistic synchrotron plasma and magnetised, thermal gas.

 Each of the aforementioned scenarios may result in a unique polarisation signal stemming from a galaxy. However, a majority of the previous studies have had limited frequency coverage and could not distinguish between the models. Narrow bandwidths greatly reduce the resolution in Faraday depth space, while broad channel widths decrease the maximum observable Faraday depth. Continuous sampling over a large frequency range allows for the recovery of the true polarisation signal.

However, the development of broad observing bandwidths in recent years has the potential to decipher contributions from individual magneto-ionic structures along the line of sight. The Compact Array Broad-band Backend (CABB, \citealt{CABB}) on the Australia Telescope Compact Array (ATCA) has opened up the a continuous frequency ranges of $1100\,-\,3100$ and $4500\,-\,6500$\,MHz, sampled at $1\,$MHz intervals. With these wide bandwidths and high spatial resolution, we are able to evaluate and model the polarisation properties of the radio emission on a pixel-by-pixel basis, investigate the origin of the polarised signal and attempt to answer whether the rotation we observe is a consequence of magnetic fields adjacent to or within the radio galaxy.

Motivated by the recent improvement in continuous frequency coverage, we observe NGC\,612 (PKS B0131-367) in an attempt to conclusively determine the physical properties of all Faraday rotating components that contributes to the observed polarisation. NGC\,612 has been studied at multiple wavelengths and is an ideal candidate with which to carry these polarisation studies. Due to its relatively low redshift ($z\,=\,0.0297$, \citealt{deVaucouleurs1991}), the two, large-scale lobes \citep{Ekers1978} span a relatively large angular scale ($\sim18$\,arcseconds) and is well-resolved by the ATCA.

Resolving NGC\,612 is particularly interesting due to the marked difference in morphology between the two radio lobes. NGC\,612 has been classified as having a hybrid radio source morphology \citep{GopalKrishna2000} with the eastern lobe exhibiting strong FR-\textsc{II} \citep{FanaroffRiley1974} characteristics, marked by a hot spot (`HS', RA(J2000) = 01:34:17, Dec(J2000) = -36:30:39), whereas the western lobe more closely matches a FR-I classification with a visible jet. The total radio power of NGC\,612  is $P_{4.8 GHz}~\sim~0.8~\times~10^{25}~$W~Hz$^{-1}$ \citep{Morganti1993}, which falls between typical FR-I and FR-II values \citep{OwenLaing1989, OwenWhite1991}.

NGC\,612 is a member of a galaxy group with 7 members \citep{Ramella2002} in which there is evidence of a recent interaction. This is supported by existence of the dust lane in the optical counterpart \citep{Ekers1978, KotanyiEkers1979} in addition to a tenuous H\textsc{i} bridge reaching 400~kpc from the disk of NGC\,612 towards its nearest neighbour, NGC 619 \citep{Emonts2008}. Furthermore, \citet{Tadhunter1993} argue that recent star-formation has taken place, as suggested by the observation of strong and narrow absorption features in the optical spectrum, with weak [O\textsc{II}] and [O\textsc{III}] emission and both \citet{Raimann2005} and \citet{Holt2007} observe a young stellar population throughout the stellar disk.

There is diffuse, soft ($0.7$\,-\,$3$\,keV) X-ray emission associated with the lobes of NGC\,612, with extended emission in the direction of the eastern lobe, described in \citet{Tashiro2000}, who argue that the emission is likely due to the cosmic microwave background up-scattering off the synchrotron-emitting electrons in the lobes. Making basic assumptions of the geometry and distribution of matter within the lobes, \citet{Tashiro2000} use the observed quantities of spectral index in both the radio and infrared regimes, as well as the surface brightness of the radio continuum and X-ray emission, to estimate an implied magnetic field strength of the lobes of $1.6\,\pm{1.3}\,\mu$G. 

NGC\,612 was selected as part of a larger sample of large angular-scale, Southern hemisphere radio galaxies, with each radio galaxy representing different intrinsic characteristics, radio morphology classifications and environments. Future work will investigate the relationship between these various radio galaxy characteristics and the observed polarisation properties. Work carried out in this paper is complimentary to recent work by \citet{Banfield2017}, who use spectro-polarimetric observations of NGC\,612 and its surrounding environment to motivate environmental impacts on radio morphology.

In this paper, we present a detailed study of the polarisation properties of the radio galaxy NGC\,612. The paper is structured as follows: we begin by describing the observations and data reduction are described in detail in $\S$\ref{n612:obs}, followed by a description of how the polarised signal was recovered from the data ($\S$\ref{n612:obsResults}). In $\S$\ref{n612:modelFitting}, we introduce the different depolarisation mechanisms and how we built and tested models of polarised emission using maximum likelihood and MCMC techniques. Our results are presented in $\S$\ref{n612:modResults}, which describes the relative success of polarisation models, in addition to introducing the parameter maps created from the best-fit polarisation solutions. The discussion ($\S$\ref{n612:discuss}) focuses on differentiating between the different depolarisation models and attempts to answer the question of the origin of the observed Faraday rotation signal. Our conclusions are presented in $\S$\ref{n612:conclude}.

\begin{figure} 
\centering
	\subfigure[$2100\,$MHz]{\includegraphics[width= \linewidth]{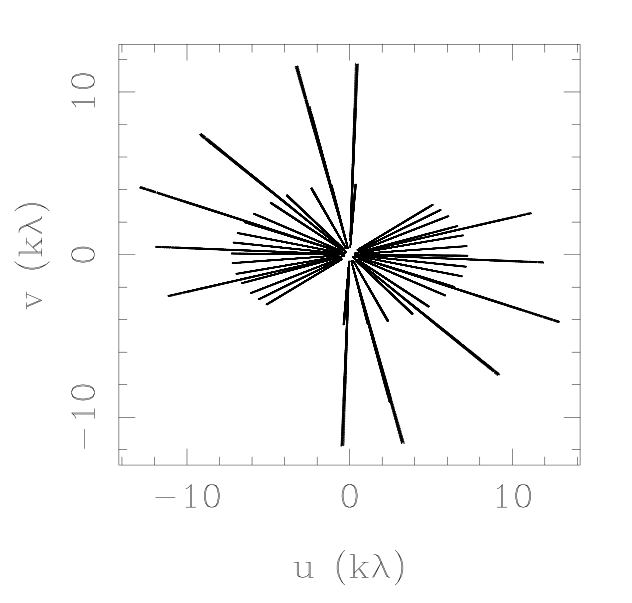}}
	\centering
	\subfigure[$5500$\,MHz]{\includegraphics[width=1\linewidth]{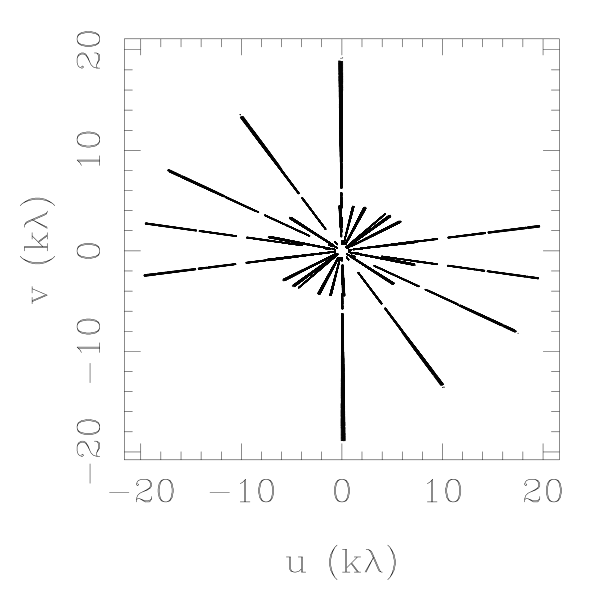}}
\caption[$uv$-coverage of NGC\,612 ]{Final $uv$-coverage for our observations of NGC\,612  for observing bands centred at $2100\,$MHz (top) and $5500$\,MHz (bottom). As the bright galaxy was observed in short snapshots, there are negligible azimuthal tracks.}
\label{n612:uvCover}
\end{figure}

\begin{figure}
\centering
 \includegraphics[width=\linewidth]{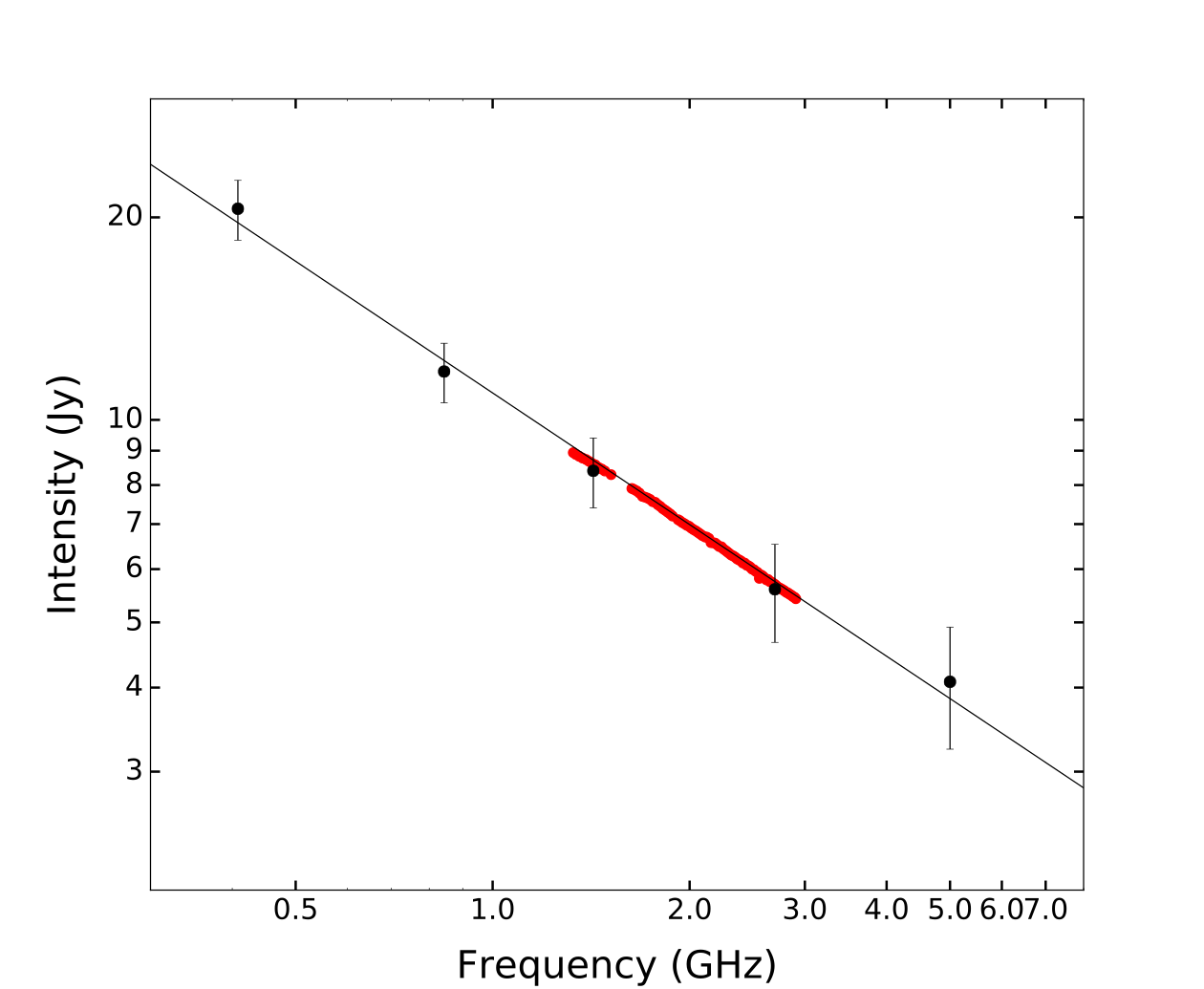}
\caption[Flux density of NGC\,612 ]{Measured flux density of NGC\,612  as a function of wavelength. Our observations are shown in red with the displayed marker size of each measured flux being larger than the measured intensity errors. Measurements taken from previous literature are in black. Reference fluxes with corresponding errors are shown at 408\,MHz \citep{SchilizziMcAdam1975}, 843\,MHz \citep{SUMSS}, 1.425\,GHz \citep{Fomalont1968}, 2.7\, GHz \citep{Bolton1973} and 5.0\,GHz \citep{Wall1979}. The figure is truncated and does not show a flux density value of 56\,Jy at 85.5\,MHz \citep{Mills1960}. Updated flux density values at 408\,MHz and 843\,GHz were acquired through private communication with Dr. Richard Hunstead. A spectral index of $\alpha$~=~-0.65 is measured from our data.  }
\label{n612:fluxDensity}
\end{figure}

\section{Observations \& Data Reduction}
\label{n612:obs}

All observations for this project were taken with the ATCA CABB \citep{CABB}, which offers 2~GHz of continuous bandwidth and a spectral resolution of 1~MHz. The wide bandwidths of CABB are ideal for recovering accurate Faraday depths from multiple contributors within the synthesised beam -- a wider range of observed wavelengths lead to better the resolving power in Faraday space \citep{BrentjensdeBruyn2005}.

 NGC\,612  was observed in multiple array configurations spanning a baseline range of $107\,-\,2923$\,m. Combining multiple array configurations allows for better sampling of the $uv$-plane leading to improved sampling of the galaxy on all angular scales. Our effective $uv$-coverage is shown in Figure~\ref{n612:uvCover} for both the observing bands covering all observing days. A summary of the observations is given in Table~\ref{n612:obsTable}, which lists the array configuration, total time on source and date of observation\footnotemark. The mosaic pointings were Nyquist sampled at the resolution of the highest frequency channel ($3100$\,MHz and $6500$\,MHz). Each mosaic pointing was observed at 30 second intervals, switching to the phase and leakage calibrator PKS\,B0153$-$410 once numerous mosaic cycles completed. The time between observations of the phase calibrator was no more than 20 minutes. 
 
 The data were reduced, calibrated and flagged using the \textsc{Miriad} software package \citep{miriad}. PKS\,B1934$-$638 was observed once per day as the absolute flux and bandpass calibrator. The bandpass, gains and polarisation solutions were calculated every 128~MHz in order to avoid any frequency-dependent calibrations. \citet{SaultCornwell1999} showed that in order to carry out polarisation calibration using an East-West array, for a source with unknown polarisation levels, observations must be made at $\geq\,3$ parallactic angles. For observations made for this study, leakage calibration was carried out using the phase calibrator source, PKS\,B0153$-$410, as observations of this target mostly resulted in sufficient parallactic angle coverage. For short observations, where multiple observations of a point source resulted in inadequate parallactic angle coverage (e.g. 2013Feb24), the bandpass calibrator source PKS\,B1934$-$638 has been used to calculate leakage solutions\footnotemark.
 
 \footnotetext{PKS\,B1934$-$638 is known to be unpolarised to less than the $0.2\%$ level. This is standard practice, as outlined in the \textsc{miriad} user's guide \citep{miriad}.}

\footnotetext{During the October 2012 and April 2013 observing semesters, there existed a large ripple running through the middle of the 5500\,MHz observing band for all observations involving ATCA antenna CA01. The ripple only affected the Y-polarisation and was highly time variant. Significant effort was made to correct the erroneous data so that we might include all baselines involving CA01; however, the attempts were ultimately unsuccessful and the high-frequency polarisation data was dropped from our analysis.}

\begin{table} 
\caption[Summary of the observing log of NGC\,612 ]{Summary of the observing log for NGC\,612, which is a subset of ATCA project C2776. Column (1) gives the array configuration; column (2) the central observing frequency in GHz. The total time spent on-source each run is listed in column (3). Column (4) gives the UT date of the commencement of the observations.} \vspace{4mm}
\footnotesize 
\centering 
\begin{tabular}{ l c c c}
Array Config. & Obs. Freq. & Time On-Source & Obs. Date  \\ 
& (MHz) & (hours) &   \\ 
\hline 
1.5 C 	 &   2100 	&	0.26 &		2012 Dec 03 \\ 
1.5 C		&	 5500	&	0.29	&		2012 Dec 03 \\
EW 352  &   5500 	&   0.77  &		2013 Jan 09 \\ 
EW 352	&   5500	&   1.37 &		2013 Jan 10\\
750 C   	& 	2100		&   0.98 &		2013 Jan 25 \\ 
750 C   	&   5500	&   0.49  &   	2013 Jan 26 \\ 
6A 	 	&   2100	&   0.1  	& 		2013 Feb 24 \\
\end{tabular}
 \label{n612:obsTable}
 \end{table}

The data were flagged largely with the automated task \textsc{pgflag}, with minor manual flagging being carried out with tasks \textsc{blflag} and \textsc{uvflag}. In total, $37\%$ and $19\%$ of the data were flagged in the 2100 and 5500\,MHz bands due to radio frequency interference (RFI), respectively.  

Naturally-weighted Stokes $I$, $Q$ and $U$ mosaic maps were made every 16~MHz. To avoid any resolution effects between frequencies, the dirty images were convolved to a common resolution of 1\,arcmin. We drop the lowest frequency maps due to insufficient $uv$-coverage after flagging.

Joint maximum entropy deconvolution was performed on the mosaics with the task \textsc{pmosmem}. Using previously published values for the total flux density of NGC\,612 , we find a spectral index value of $\alpha\,=\,-0.65$ ($S\,\propto\,\nu^{\alpha}$). In the absence of single dish observations for an absolute flux reference, we estimate the expected total source flux extrapolating from this spectral index value and previous measurements (Figure~\ref{n612:fluxDensity}). To test the validity of this assumption, we additionally cleaned the Stokes $I$ maps with a {\em multiscale clean} approach with the {\em Common Astronomy Software Applications ({\sc CASA})}. We find that corresponding frequency Stokes $I$ maps are nearly identical, with negligible variations in both total flux and on a pixel-by-pixel basis compared to the rms noise. 

Cleaned images were generated with the task \textsc{restor}. Our synthesised beam is $1\,$arcmin across where one pixel corresponds to 10~arcsec in angular size. At the distance to the galaxy of 121.5~Mpc, one beam corresponds to a physical size of $\sim35$~kpc for $H_{0}\,=\,73.0$\,km/sec/Mpc.

\section{Observational Results}
\label{n612:obsResults}

\subsection{Imaging Results}
\label{n612:imageResults}

\begin{figure*}
\centering
\includegraphics[width=1\linewidth]{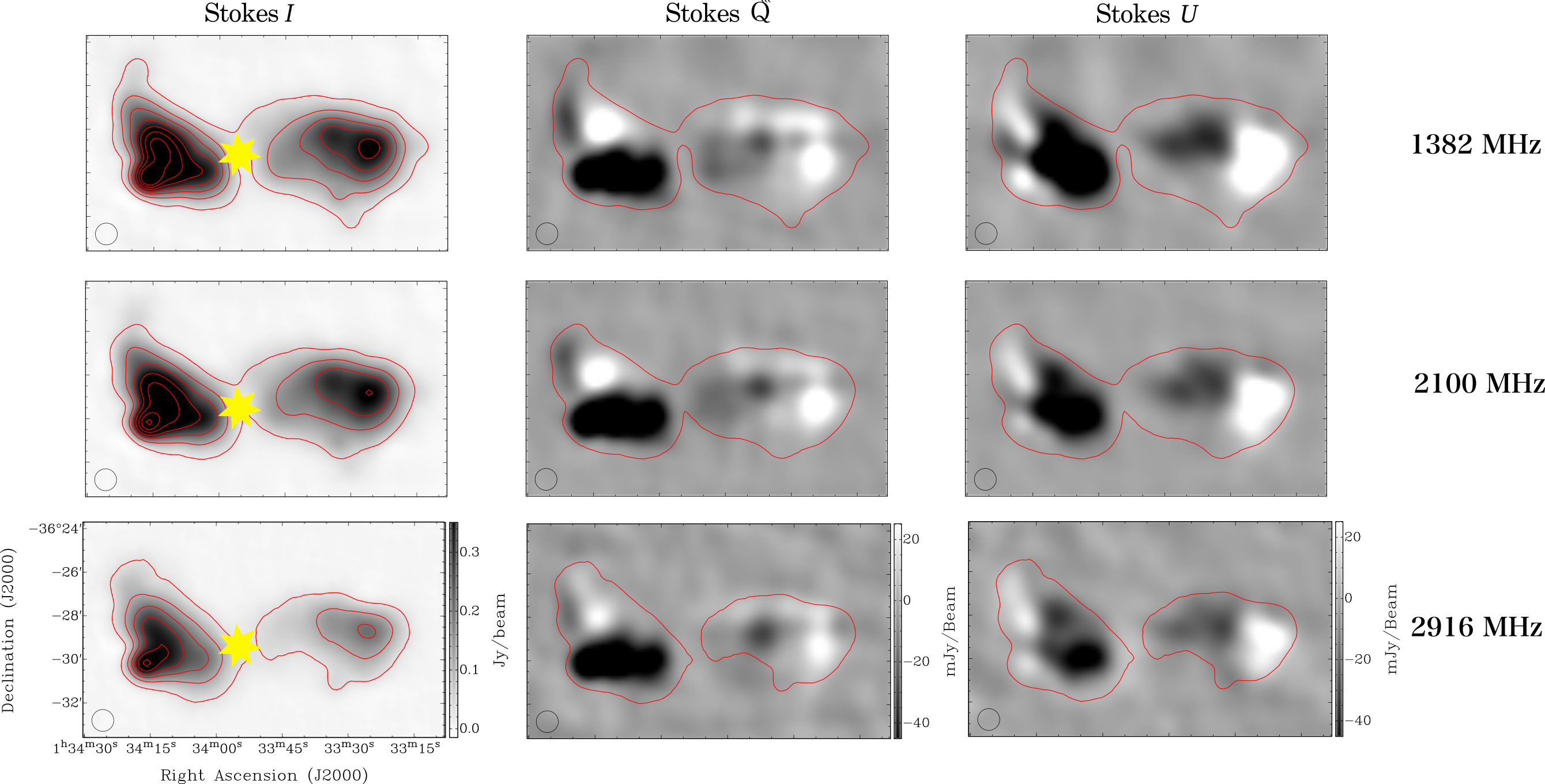}
\caption[Frequency-dependent Stokes maps of NGC\,612 .]{Maps of Stokes $I, Q$ and $U$ (left, centre and right respectively) at wavelengths centred at 1332, 2100 and 2916\,MHz in ascending order. These frequency channels represent the extremes and middle of the $2100\,$MHz observing band. Intensity levels are kept constant across the displayed frequency range with the colour bar shown in the bottom panel. The circular beam has a FWHM of 1\,arcminute and is shown in the lower left-hand corner of each map. Stokes $I$ contours are shown in red and represent total intensity values of $30$ - $480$ mJy/beam at intervals of $75$ mJy/beam for the total intensity maps and the lowest intensity contour is shown on both the Stokes $Q$ and $U$ maps. The Stokes $I$ rms noise levels are $\sigma_{I}\,=\,5.4, 3.7$ and $4.1$\,mJy\,beam$^{-1}$ for 1382\,MHz, 2100\,MHz and 2916\,MHz, respectively. The rms noise levels of the Stokes $Q$ and $U$ maps are equivalent at each frequency with values of $\sigma_{Q,U}\,=\,2.0, 1.5$ and $1.9$\,mJy\,beam$^{-1}$ for 1382\,MHz, 2100\,MHz and 2916\,MHz.} The location of the optical galaxy is marked with a yellow star in all total intensity figures.
\label{n612:stokesMaps}
\end{figure*}

Our data reduction resulted in 92 independent channel maps in Stokes $I$, $Q$ and $U$. Typical maps of total intensity (Stokes $I$) and both linear polarisations (Stokes $Q$ and $U$) are shown every 512~MHz in Figure~\ref{n612:stokesMaps}. We see a signal with clear linearly polarised emission stemming from both lobes. Corresponding frequency maps of polarised intensity ($P~=~\sqrt{Q^{2}+U^{2}}$) were created from the final Stokes maps. It is immediately evident that the polarisation signal changes as a function of position across NGC\,612  (Figure~\ref{n612:stokesMaps}).

Uncertainties in the intensity values of $I, Q, U$ and $P$ were measured for each frequency interval by taking the rms value ($\sigma$) of an area in the final maps near, but not including, the radio emission. Pixels were masked if $>10\%$ of the channels fell below a threshold of $8\sigma$ in Stokes $I$ and $5\sigma$ in polarised intensity. We discard all edge pixels from the continuous, accepted pixels that comprise the radio galaxy. This results in 1,277 usable pixels comprised of 45 independent beams for our analysis.

\begin{figure*}
\centering
\includegraphics[width=\textwidth]{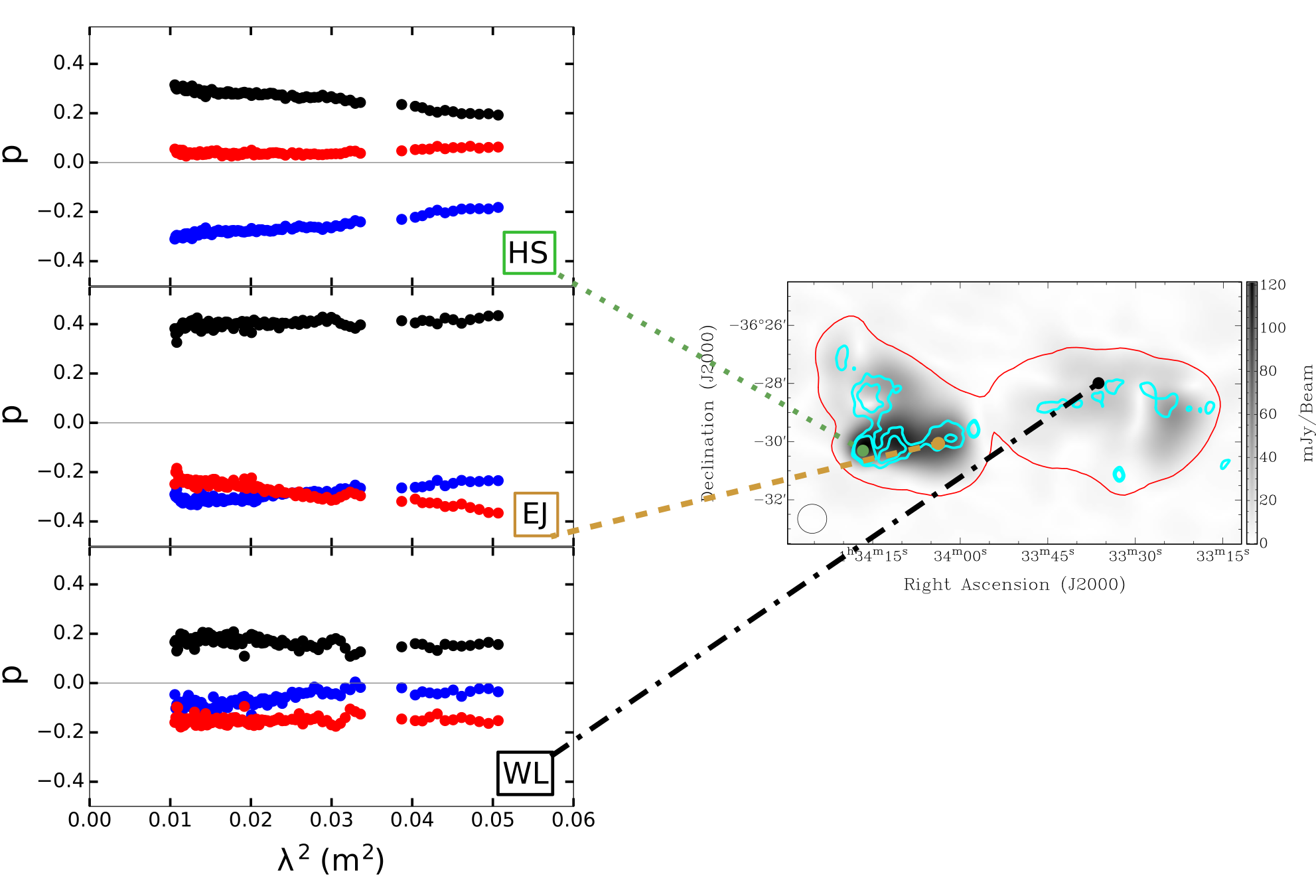}
\hfill
\caption[Example fractional polarisation spectra from NGC\,612 .]{Polarised intensity map at 2100~MHz (\textit{right}) with red contours outlining 30 mJy/beam in total intensity (Stokes $I$). Fractional polarised spectra are shown for the indicated pixel locations. Polarised fraction, $p$ ($P/I$) is shown in black, $q$ ($Q/I$) in blue and $u$ ($U/I$) in red versus $\lambda^{2}$. All data points are displayed on a scale larger than their corresponding errors. Point `HS' represents the location of the hot spot in total intensity, point `EJ' (eastern jet) is near the location of the optical counterpart (identified in Figure~\ref{n612:stokesMaps}) and point `WL' is a pixel in the western lobe. The path of the jet, as determined from high-frequency imaging, is shown in cyan contours.}
\label{n612:polSpectra}
\end{figure*}

We also make a high-resolution, high-frequency total intensity map at 5500\,MHz in order to trace the path of the jet associated with the lobes of NGC\,612 . The Stokes $I$ imaging results are minimally effected by the hardware issues described in $\S$\ref{n612:obs}, leading the deconvolved map of Stokes I to be robust and giving us confidence in the jet position. Contours of the high-frequency position of the jet is marked by the cyan contours in Figure~\ref{n612:polSpectra} and as a grey dashed line in all subsequent parameter maps of the radio lobes. 

\subsection{Fractional Polarisation Spectra}
\label{n612:fracSpec}

In order to decouple spectral effects from wavelength-dependent polarisation, we adopt fractional polarised notation, such that $q = Q/I$, $u = U/I$ and $p^2 = q^2 + u^2$. We create fractional polarised spectra, by dividing the observed $Q(\lambda^2)$ and $U(\lambda^2$) by a second-order polynomial model, bootstrapped to the Stokes $I$ emission. We fit the Stokes $I$ spectrum in linear $S(I)$ versus $\nu$ space to avoid over-weighting higher frequency flux density measurements, thus leading to the creation of non-Gaussian noise when propagating the uncertainty as a function of frequency.

Figure~\ref{n612:polSpectra} shows the fractional polarisation spectra of a few representative pixels across the source and also demonstrates the varying levels of polarisation seen in the galaxy. Depolarisation is defined as a negative change in the observed degree of polarisation as a function of $\lambda^2$ ($dp/d\lambda^2\,<\,0$). We observe this trend most clearly in the region of the hot spot (`HS'), where observed level of polarisation decreases as a function of $\lambda^2$. 

\begin{figure}
 \subfigure[]{\includegraphics[width=\linewidth]{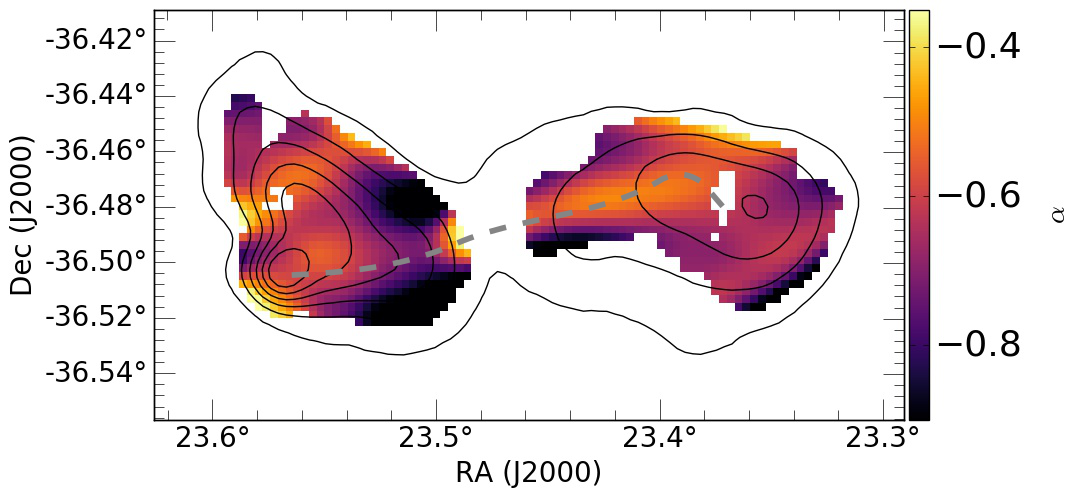}}
	\centering
  \subfigure[]{\includegraphics[width=1\linewidth]{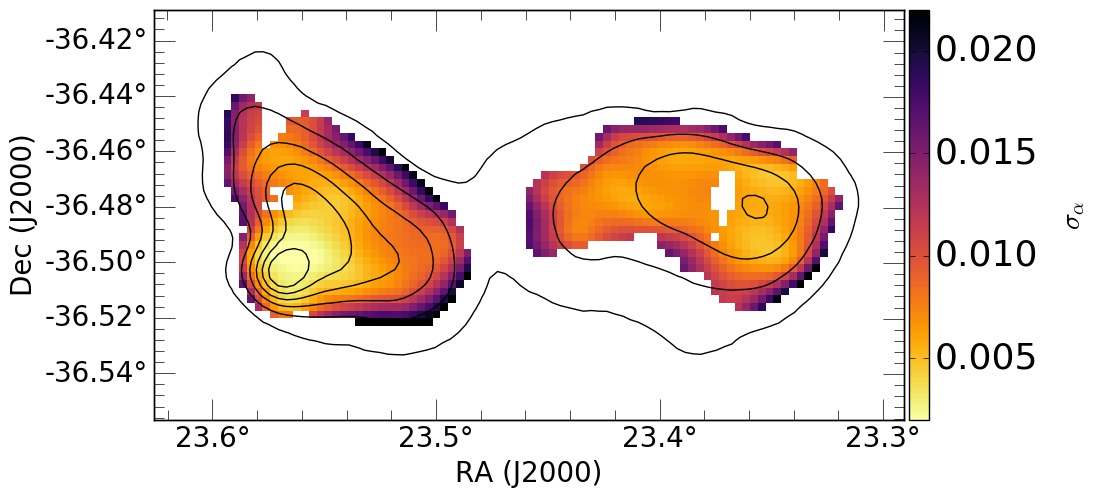}}
\caption[Spectral index map of NGC\,612.]{Spectral index values for each pixel evaluated as a single-component linear fit to the Stokes $I$ spectrum in log-space (a) with corresponding 1$\sigma$ uncertainty (b). In (a), the dashed grey line shows the path of the jet. Contours show the total intensity from 25 - 400 mJy/beam in intervals of 75 mJy/beam.}
\label{n612:specIndex}
\end{figure}

Assuming the lobes are composed of an optically thin synchrotron radiation source, we calculate the spectral index for each extracted pixel by fitting the Stokes $I$ spectrum to a single, emission component in log-space. In making this assumption we are also assuming that the dominant polarisation component is also the dominant spectral component, and this may not necessarily be true in the case of the projected area of the jet where it is possible to have two significant spectral components. Figure~\ref{n612:specIndex} shows a map of spectral index across the lobes of NGC\,612 . The dashed line traces the projected path of the jet through the lobes. 

 Our analysis ($\S$\ref{n612:modResults}) will partially focus on the three representative pixels shown in Figure~\ref{n612:polSpectra}. The hot spot, denoted `HS', represents the area of bright continuum emission at the far east end of the eastern lobe. The eastern jet, denoted `EJ', is a pixel that is near the optical galaxy and marks the position of peak polarisation. This pixel is located within the jet that can be seen stretching across the eastern lobe. The pixel in the western lobe, denoted `WL', is a representative pixel for the majority of this lobe, which has on average a lower degree of polarisation than the eastern lobe.

\section{Model Fitting}
\label{n612:modelFitting}

In highly energetic environments such as radio lobes, complex magnetic field and thermal electron structures will give rise to multiple rotation and/or emitting sources along the line of sight and the observed polarisation signal may experience depolarisation. Depolarisation may be a result of turbulent mixing of gas in the lobes, the emitting and rotating material being co-spatial, or the spatial resolution of the observations. The effects of depolarisation tend to be strongest towards longer wavelengths and the mechanism responsible contains insight on the overall structure of the radio lobes. 

 The observation of depolarisation in Figure~\ref{n612:fracSpec} at the location of the hot spot makes it immediately evident that it will be necessary to involve some polarisation models that are more complex (i.e. not Faraday thin) in order to accurately describe at least some of the observed polarisation. In order to analyse the nature of the polarisation of NGC\,612  and its surrounding area, we test various single-component polarisation models, which we detail below . 

\subsection{Polarisation Models}
\label{n612:models}

Polarised emission in the absence of a Faraday rotating medium can be expressed simply as
\begin{equation}
\mathcal{P} = p_{0} e^{2i\Psi_0}
\end{equation} 
\noindent where the $p_{0}$ is the intrinsic fractional polarisation, $\Psi_{0}$ is the intrinsic polarisation angle of the emission. If the polarised emission passes through a purely foreground, the polarisation angle is rotated from its intrinsic value. This is the simplest scenario of Faraday rotation and the polarised signal that is observed can be expressed as
\begin{equation}
\mathcal{P} = p_{0} e^{2i(\Psi_{0} + \phi~\lambda^{2})},
\end{equation}
where $\phi$ is the Faraday depth at the distance of the magneto-ionised material. However, the observation of depolarisation in Figure\,\ref{n612:polSpectra} at the location of the hot spot makes it immediately evident that it will be necessary to involve more complex models in order to accurately describe at least some of the observed polarisation. 

There are typically three mechanisms that can lead to depolarisation, all of which are presented with complete discussion in \citet{Sokoloff1998} and references therein, and all we briefly outline below.

\vspace{2mm}
\subsubsection{External Faraday Dispersion (EFD, Beam Depolarisation)}
If many turbulent polarising cells are within the telescope beam -- or if there is an ordered magnetic field that varies in strength and/or direction -- the polarised emission will undergo different amounts of rotation along the different lines of sight. When the fluctuations are averaged across the entire beam area, the result is depolarisation of the form 
\begin{equation}
\mathcal{P}~=~p_0~e^{2i(\Psi_0+\phi\lambda^2)}~e^{-2\sigma^{2}_{\phi}\lambda^4},
\end{equation}
\noindent where $\sigma_{\phi}$ characterises changes in Faraday depth on scales smaller than our beam. This type of depolarisation may occur when an expanding radio lobe sweeps up material from its surrounding environment. If this material is ionised or threaded by a homogeneous magnetic field, it would act as a thin Faraday screen, but would not depolarise the emission from the radio lobes \citep{Bicknell1990}.  Fluctuations in the magnetised intergalactic medium is the other, albeit more common, application of external Faraday dispersion and has been used to study magnetic fields in galaxy groups and clusters \citep{Laing2008_structures}.

Due to the purely external dependence of this type of depolarisation, and its dependence on the observing beam, this type of depolarisation is often referred to as \textit{`beam depolarisation'}. \\

\subsubsection{Differential Faraday Rotation (DFR, Depth depolarisation)}
 If the emitting and rotating medium are mixed in the presence of a regular magnetic field, the emission on the far side of the radio lobe will experience a different amount of rotation when compared to the emission at the near of the lobe. When summed over the line of sight, this will lead to depolarisation that takes the form of 
\begin{equation}
\mathcal{P}~=~p_{0}~e^{2i(\Psi_{0}+\frac{1}{2}~\phi\lambda^{2}})~\,\frac{\sin~\phi\lambda^{2}}{\phi\lambda^{2}}
\end{equation}
\noindent for a symmetric, uniform slab, where $\phi$ is the Faraday depth through the region. For an emitting and rotating medium of an arbitrary thickness, the classical RM is equal to $\frac{1}{2}\phi$, i.e. the actual Faraday depth is equal to twice the observed RM. \\

\subsubsection{Internal Faraday Dispersion (IFD)}
This scenario is similar to DFR, but rather than being concurrent solely with a uniform magnetic field, the emitting and rotating medium may also be in the presence of a turbulent magnetic field. Any emission now must undergo a random walk through the turbulent field, which results in the plane of polarisation being rotated different amounts. For the simplest case, we consider a Gaussian distribution of Faraday depths within the lobes,
\begin{equation}
\mathcal{P}~=~p_{0}~e^{2i\Psi_{0}} \frac{1~-~e^{2i\phi\lambda^{2}-2\zeta^{2}\lambda^{4}}}{{2\zeta^{2}\lambda^{4}-2i\phi\lambda^{2}}} ,
\end{equation}
\noindent with a mean Faraday depth $\phi$ and standard deviation $\zeta$. IFD characterises both the uniform and turbulent magnetic fields, with the behaviour becoming identical to that of DFR in the case of a dominating uniform field.\\

\subsubsection{Multi-Component Emission}
There is also the possibility that there are multiple emitting, rotating and/or depolarising components along the line of sight.  Multiple component models can be constructed by summing the complex polarisation from individual models
\begin{equation}
\noindent\mathcal{P}~=~\mathcal{P}_{1}+\mathcal{P}_{2}+\mathcal{P}_{3}+ ... +\mathcal{P}_{N}.
\end{equation}
\noindent  A complex polarisation signal can often have a beating pattern as a function of $\lambda^{2}$ as the rotation due to different components passes into and out of phase.

We limit our investigation to a single polarised component. Upon inspection, the higher order features of the $q$ and $u$ spectra are likely contaminated by instrumental effects. This minimises the possibility of over-fitting the data. In any case, one component is sufficient to describe the bulk of the polarised signal. This limits the physical interpretation of the polarised morphology, but as we will demonstrate in the sections that follow, acceptable solutions can be found. We found that expanding the number of components to be greater than one can fit the data closely; however, the degeneracy between two-component models and their parameters prevents interpretation.

 \subsection{Modeling Procedure}
\label{n612:quFitting}

 \begin{figure*}
	\includegraphics[width = 0.48\linewidth]{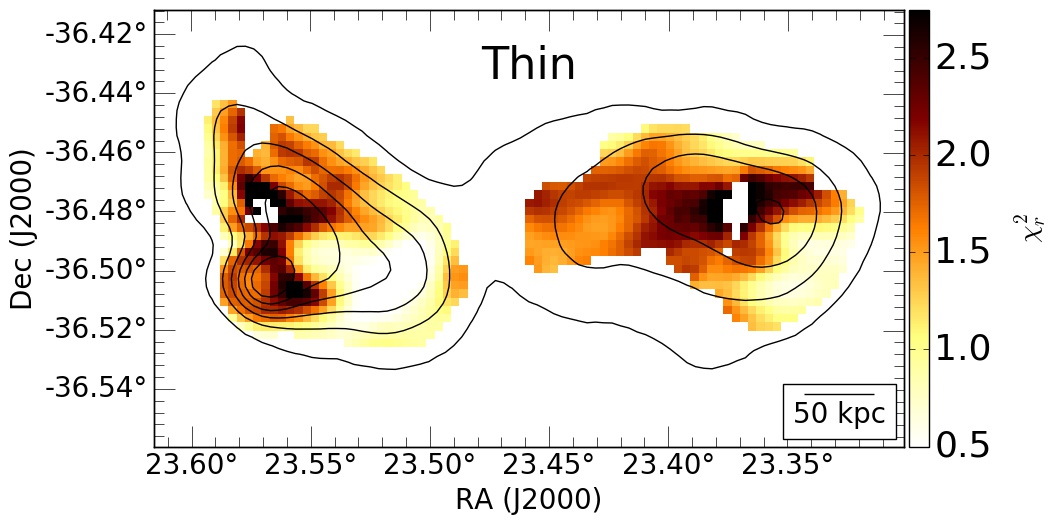}
	\hfill
	\includegraphics[width = 0.48\linewidth]{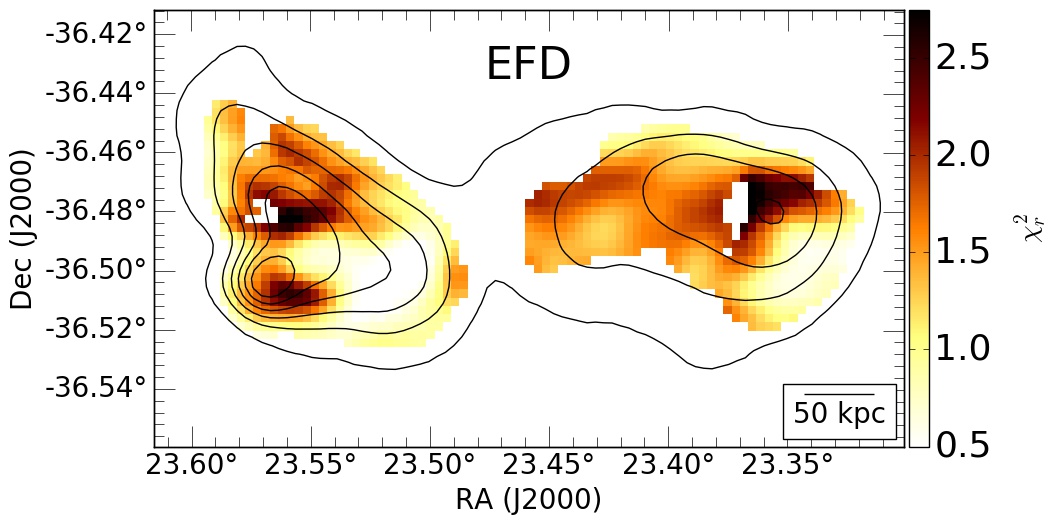}
	\\[6mm]
		\includegraphics[width = 0.48\linewidth]{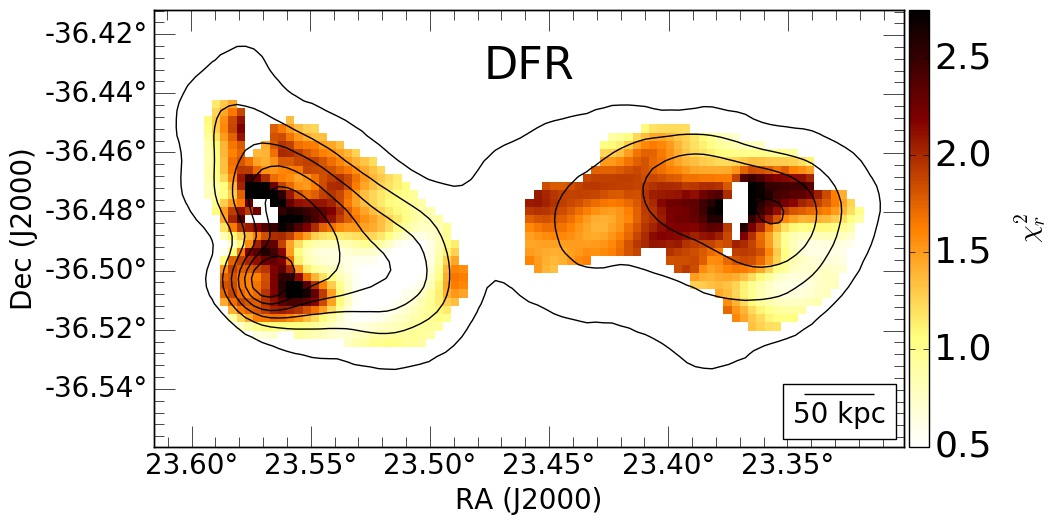}
	\hfill
	\includegraphics[width = 0.48\linewidth]{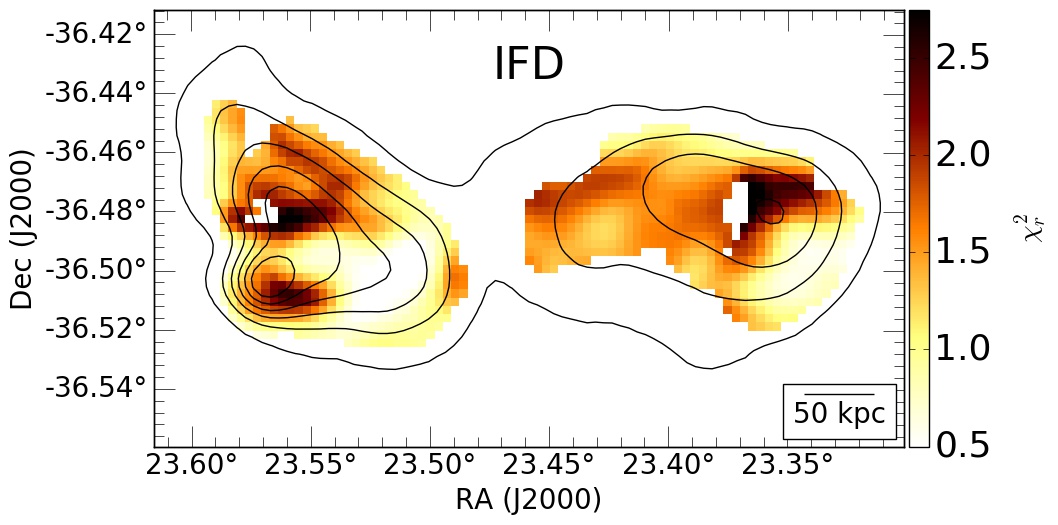}
\caption[$\chi_r^2$ maps returned from the tested polarisation models.]{$\chi_r^2$ values at each imaged pixel for all polarisation models discussed in this work. Black contours outline total intensity levels spanning $25\,-\,400$\,mJy/beam in $75$\,mJy/beam increments. A scale bar is shown in the bottom righthand corner of each figure. All models have a tendency to have similar $\chi^2$~values.}
\label{n612:chi2} 
\end{figure*}

 In order to test which mechanisms may be responsible for the observed polarisation, the single-component depolarisation models introduced in $\S$\ref{n612:models} (i.e. thin, EFD, DFR, IFD) are explored individually. The extracted $q(\lambda^{2})$ and $u(\lambda^2)$ data are simultaneously fit to each corresponding depolarisation model using a maximum likelihood method. We employ the EMCEE \textit{Python} module \citep{emcee} to fit the complex polarisation parameters of each depolarisation model. Unlike Levenburg-Marquardt fitting, MCMC has the added benefit of better exploration of parameter space, as well as returning numerically determined errors to model parameters. The log-likelihood of the complex polarisation model of the joint $q?,? u$ chi-squared ($\chi^{2}$) is minimised to find the best-fitting parameters. For each pixel in our dataset, we initialise a set of 250 parallel samplers that explore the $n$-dimensional parameters space (where $n$ is the degrees of freedom). Each of these walkers iteratively calculate the likelihood of a given location in parameter space and in doing so map out a probability distribution for a set of parameters. For any given depolarisation model, the possible parameter solutions were forced to be physical (i.e. $0\leq~p~\leq\,1$). In order to maximise the effectiveness of initial burn-in phase and have the walkers settle on a parameter space, each iteration of MCMC used the previous pixel's best-fit values as an initial guess. 
 
 To assess the goodness-of-fit of each models, the reduced chi-squared value ($\chi^2$ ) was recorded for each pixel. $\chi^2$  is defined as
\begin{equation}
\chi^2_r = \frac{1}{\nu}\sum\limits_{i\,=\,1}\limits^{n}\left(\frac{x_i\,-\mu_i}{\sigma_i}\right)^2,
\label{eq:chi2}	
\end{equation}
\noindent where $x_i$ is the $i^{\rm th}$ independent data point with Gaussian noise, $\sigma_i$; $\mu_i$ is the model prediction and $\nu$ is the number of free parameters. This statistic penalises according to how many standard deviations lie between the observed point and the model and generally serves as a means of assessing the success of a particular model fit to the observed data. However, it can be misleading to employ $\chi^2$  when comparing the relative success between models, as it is possible to build, and fit, arbitrarily complex models.

In the case of Gaussian noise, the $\chi^2$ is equivalent to $-2$log$\mathcal{L}$. We note that $\chi^2$ is not an ideal statistic with which to evaluate the success of these models as our sampling of $\lambda^2$-space is non-linear, which in-turn overweights the importance of low-$\lambda^2$~samples and will down-weight samples at high-$\lambda^2$. Additionally, for a sufficiently robust dataset of a simple polarised spectrum, a returned $\chi^2$ $<1$ is not uncommon. This is due to the minimum number of model parameters of any physical polarisation model being 3 ($p_0, \Psi_0, \phi$) and a model with fewer parameters is deemed unphysical.  In these instances, we acknowledge the over-fitted solution, but use the returned best-fit parameters for our analysis.

\section{Modelling Results}
\label{n612:modResults}

{Fitting a single polarised component to NGC\,612  returns a $\chi^2$ $\leq\,1.5$ for more than $60\%$ of pixels (Figure~\ref{n612:chi2}); expanding this to include all pixels with $\chi^2$ $\leq\,2$ results in the acceptance of more than $90\%$ of pixels across the lobes of NGC\,612 . We therefore believe that one polarisation component is sufficient to describe the bulk of the polarised signal. Upon inspection, any higher order features of the $q$ and $u$ spectra are likely contaminated by instrumental issues that affected the ATCA at the time of observation (e.g. 300\,MHz spectral wiggle, $\S$\ref{n612:obs}). Therefore,  we limit our investigation to a single polarised component.} This minimises the possibility of over-fitting the data and over-interpreting the results. Although this limits the physical interpretation of the polarised morphology, we will demonstrate in the sections that follow that a favourable solution can be found by examining the physics of the implied model. In the following section we explore the relative success of each model with the aim of determining which best represents the polarisation signature associated with the radio galaxy.  
 
 Figure~\ref{n612:modelComp} represents the best-fit solutions to each of the polarisation models for three single-pixel spectra. The pixels are selected to be representative of three independent regions of the radio lobes: the hot spot, the east jet and the western lobe (see Figure~\ref{n612:polSpectra} for specific locations). Below each model fit, the residual polarisation spectrum is shown to demonstrate any latent structure in the spectrum. The similarity between each model's best-fit is immediately evident, with the largest discrepancy between model solutions occurring at the location of the hot spot ({\em top row}, Figure~\ref{n612:modelComp}). In this region, only polarisation mechanisms with an explicit dispersion term (EFD, IFD) are able to fit the spectra at large $\lambda^2$.

 Global $\chi_r^2$ values for each polarisation model are shown in Figure~\ref{n612:chi2}. The $\chi^{2}_{r}$ maps allow the reader to assess a model's overall success in fitting the polarisation signal of the entire radio galaxy. Table~\ref{n612:chi2Table} reports the mean reduced chi-squared ($\overline{\chi}^2_r$ ) for an area equivalent to the synthesised beam in three regions of the radio lobes. Given the relative similarity in model-fitting results, as shown in Figure~\ref{n612:modelComp}, it is unsurprising that each polarisation model also returns a similar global success. In addition to the region of the hot spot, the western lobe has a marginal preference for dispersion models (see Table~\ref{n612:chi2Table}). 

\begin{table}
\centering
\caption[Table of $\chi^2_r$ values for a selection of points in NGC\,612 .]{Mean $\chi^2_{r}$ for each depolarisation models discussed in $\S$\ref{n612:modelFitting} for the locations shown in Figure \ref{n612:polSpectra}. The mean is calculated for a number of pixels equivalent to our synthesised beam. }
\begin{tabular}{lcccc}
Pixel Location 	& 	Thin		&		EFD		& 	DFR		&		IFD 	\\
\hline
Hot Spot 				& 		1.9			&			1.7			& 		1.9			&			1.7			\\
Eastern Jet			&			0.40 		&			0.40		&			0.42 		&			0.43		\\
Western Lobe 	& 		2.0			&			1.6			&			1.9			&			1.6			\\
\end{tabular}
\label{n612:chi2Table}
\end{table}

\subsection{Parameter Maps}
\label{n612:paramMaps}

In this section, we present the best-fit parameter maps returned from our $qu$-fitting routine. The parameter values for each of the polarisation models are similar enough to allow us to present the general signal trends here. Detailed parameter maps and their corresponding uncertainties for each individual polarisation model as given in the Appendix.

\subsubsection{Intrinsic Degree of Polarisation}
\begin{figure}
\includegraphics[width=1.\linewidth]{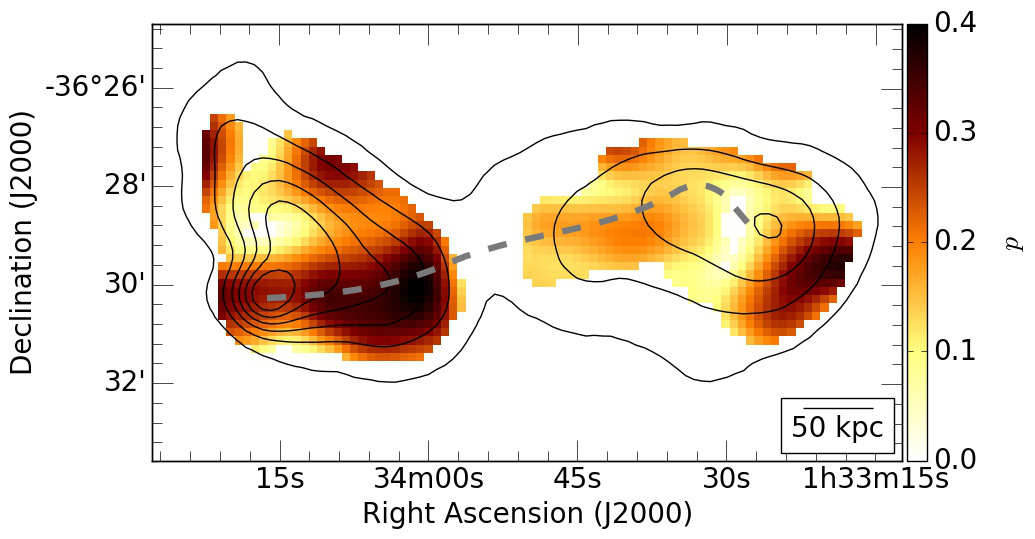}	
\caption[Map of the intrinsic polarisation fraction across NGC\,612 .]{Map of the intrinsic polarisation fraction in the radio lobes. All polarisation models yield similar intrinsic polarisation maps. The grey dashed line traces the peak in polarised intensity, which we believe to be representative of the path of the jet. Black contours outline total intensity levels spanning $25\,-\,400$\,mJy/beam in $75$\,mJy/beam increments.}
\label{n612:p0}
\end{figure}

Figure~\ref{n612:p0} shows the intrinsic polarisation across the lobes of NGC\,612 . The polarisation signal peaks nearest the location of the optical galaxy (shown as a yellow star in Figure~\ref{n612:stokesMaps}). There appears to be strong polarisation along the path of the jet in the eastern lobe through to the hot spot whereas the polarisation in the western lobe peaks at the edge of the lobe furthest away from the optical counterpart. The grey dashed line shown in Figure~\ref{n612:p0} traces the jet, as identified from high-frequency observations (Figure \ref{n612:polSpectra}).

\subsubsection{Intrinsic Polarisation Angle}
\begin{figure}
\centering
	\includegraphics[width=1\linewidth]{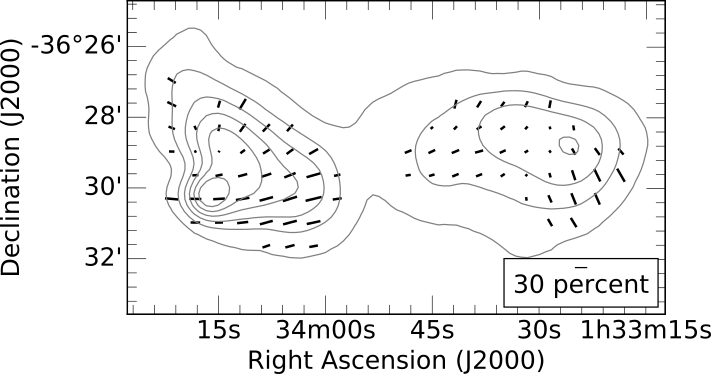}
	\caption[Intrinsic polarisation angle across NGC\,612 .]{Map of the intrinsic polarisation angle ($\Psi_0$) towards NGC\,612 . The length of each vector is proportional to the polarised intensity, with a representative fraction represented in the bottom left corner. Black contours outline total intensity levels spanning $25\,-\,400$\,mJy/beam in $75$\,mJy/beam increments.}
\label{n612:psiMaps}
\end{figure}

Figure~\ref{n612:psiMaps} presents the intrinsic polarisation angle of the electric field vector ($\Psi_0$). The orientation of the plane-of-the-sky magnetic field ($B_\perp$) is orthogonal to the position angle of the electric vector. The length of each vector in Figure \ref{n612:psiMaps} is representative of the relative degree of polarisation, with the longest vector equivalent to 40\% polarisation. We see coherence in the direction of the intrinsic polarisation angle on scales larger than the scale of our synthesised beam. In both the eastern and western lobe, the direction of the polarisation angle nearest the optical counterpart appears to be nearly parallel to the direction of jet launch, which X-ray observations have found to be nearly perpendicular to our line of sight \citep{Eguchi2011}.

 \begin{landscape}
\centering
\vspace{10mm}
\begin{figure}
\includegraphics[width = 1.33\textwidth]{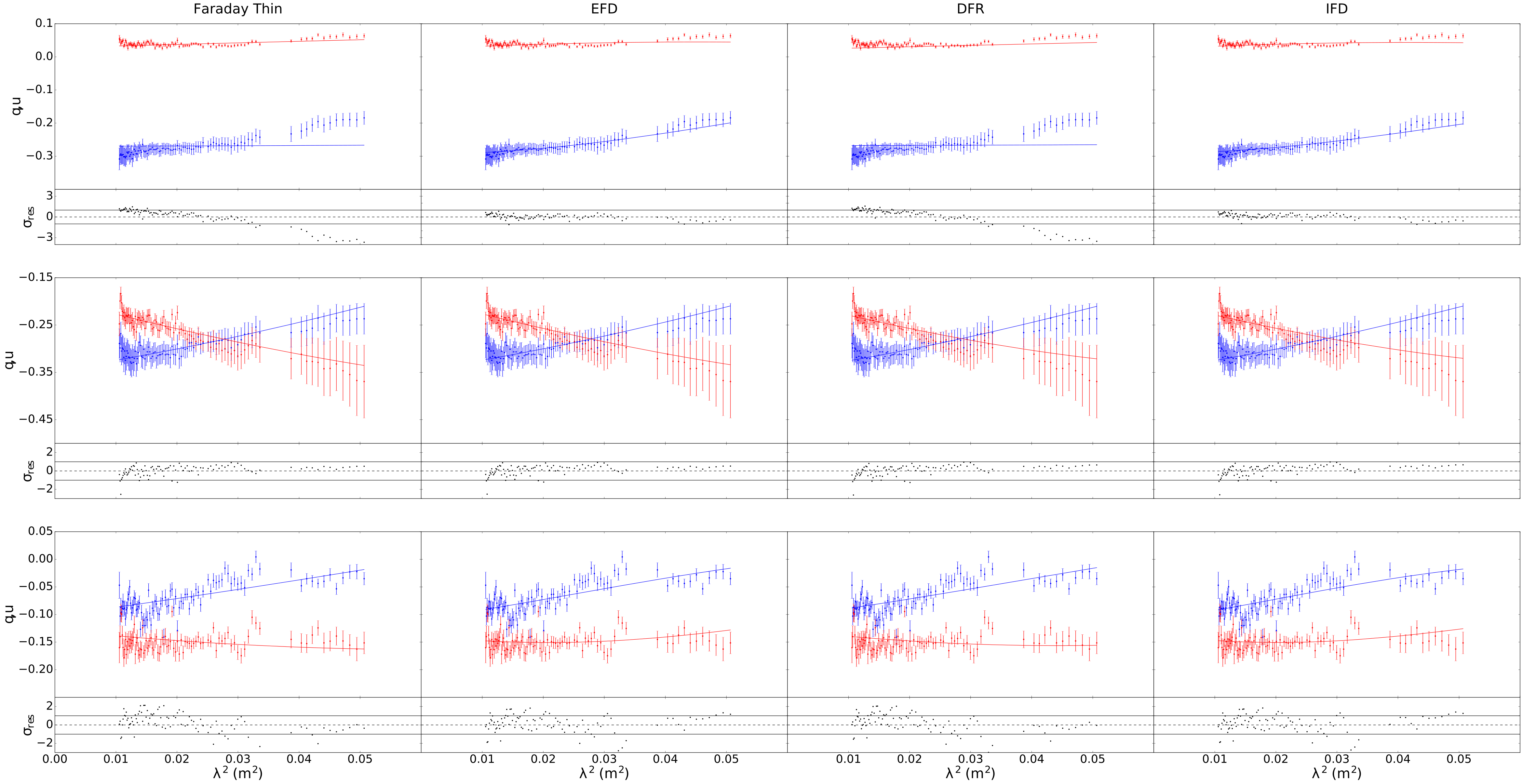}
\caption[Examples of $qu$-fitting results for the tested polarisation models.]{Grid of best-fit results from $qu$-fitting to three distinct pixels in NGC\,612 . The columns represent the models tested, while each row represents the three pixels tested. Each pixel was chosen from a unique region in the radio lobes -- the hot spot ({\em top}), the eastern jet ({\em middle}) and western lobe ({\em bottom}). The blue and red points are the measured $q$ and $u$ data, respectively. The blue and red lines are the resulting joint best-fits from $qu$-fitting. The bottom panels of each cell show the residual degree of polarisation ($(p_{obs}-p_{mod})/\sigma_{p}$) for each model, with the solid black lines indicating +1 and -1$\sigma_{p}$ deviations from the observations. }
\label{n612:modelComp}
\end{figure}
\end{landscape}

\begin{figure}
	\includegraphics[width=1.\linewidth]{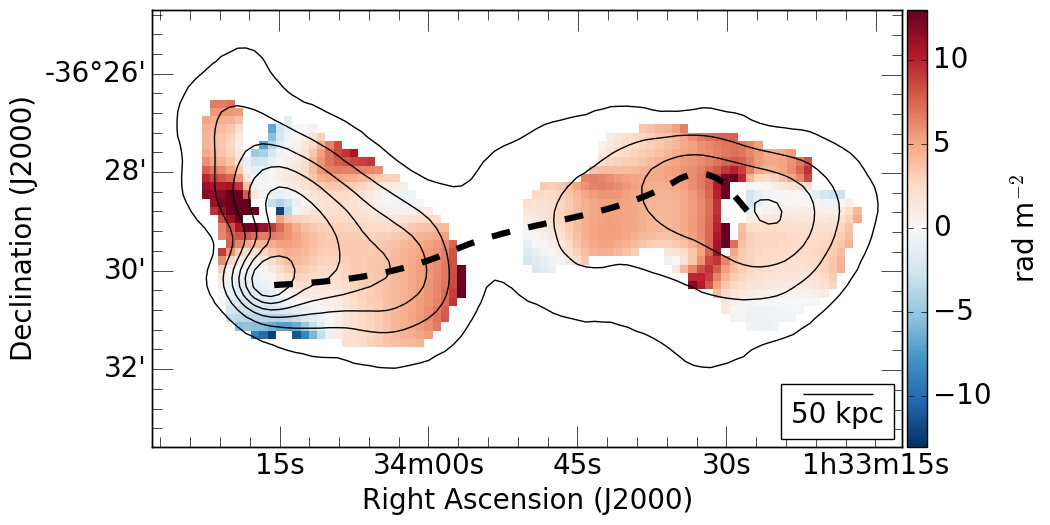}
	\caption[Faraday depth across NGC\,612.]{Faraday depth across NGC\,612 , as returned from the EFD model. The grey dashed line traces the path of the jet. DFR and IFD models show a similar trend but return $\phi$-values that are a factor of 2 larger in magnitude per pixel. Black contours outline total intensity levels spanning $25\,-\,400$\,mJy/beam in $75$\,mJy/beam increments.}
	\label{n612:rmMaps}
\end{figure}

\subsubsection{Faraday Depth}
Maps of Faraday depth are shown in Figure~\ref{n612:rmMaps}. The signal is nearly homogeneous across the source and varies on scales that are generally many times larger than our 1 arcminute beam. The dominant Faraday depth signal across the radio lobes of NGC\,612  is positive, implying that the magnetic field along the line-of-sight is oriented predominantly towards the observer. The only considerable exceptions to this orientation is surrounding the location of the hot spot in the eastern lobe, where the Faraday depth sign is predominantly negative. 

DFR and IFD models return a similar trend to that seen in Figure~\ref{n612:rmMaps} but $\phi$-values are a factor of 2 larger per pixel. As we have discussed in $\S$\ref{n612:models}, this is an expected result. The mean Faraday depth for the Faraday simple components (thin, EFD) are $\overline{\phi}_s\,=\,+3.3\,\pm\,3.5$\,rad\,m$^{-2}$.  The mean Faraday depth of the complex models (DFR, IFD) is $\overline{\phi}_c\,=\,+6.8\,\pm\,6.7$\,rad\,m$^{-2}$. Previous Faraday rotation studies of the lobes of NGC\,612  have found a simple Faraday depth that is in agreement with both $\overline{\phi}$ estimates ($+7\,\pm\,5$\,rad\,m$ ^{-2}$, \citealt{Haves1975}; $+6\,\pm\,1$\,rad~m$^{-2}$, \citealt{Simard-Normandin1981}).

\subsubsection{Faraday Dispersion/Depolarisation}

\begin{figure}
	\includegraphics[width=1.\linewidth]{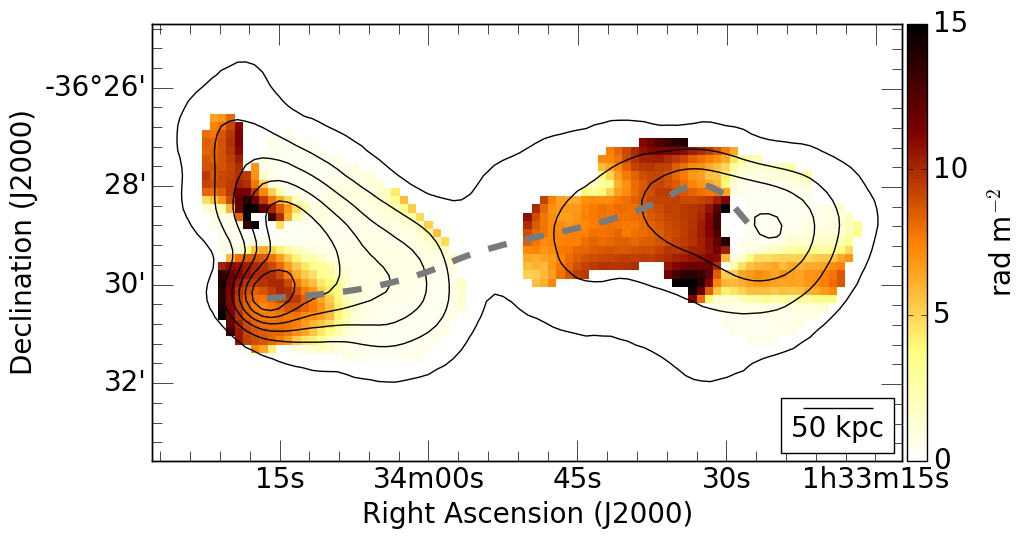}
	\caption[Map of Faraday dispersion towards NGC\,612 .]{Map of Faraday dispersion ($\sigma_{\phi}$) returned from $qu$-fitting. Best-fit values of internal Faraday dispersion ($\zeta$) are double that of the external Faraday dispersion. Black contours outline total intensity levels spanning $25\,-\,400$\,mJy in $75$\,mJy increments. The grey dashed line traces the path of the jet, as identified in higher-frequency observations.}
	\label{n612:depolMaps}
\end{figure}

Figure~\ref{n612:depolMaps} shows values of external Faraday dispersion ($\sigma_{\phi}$) for all pixels across the radio lobes. By contrast, the internal Faraday dispersion ($\zeta$) values returned are twice the magnitude of those in Figure~\ref{n612:depolMaps}. Assuming that depolarisation scales with dispersion levels, we see varying level of depolarisation as a function of position on the lobes of NGC\,612 .

Comparing similar $\chi^2_r$ values from Figure~\ref{n612:chi2}, we see that although models with dispersion terms (e.g. EFD and IFD) yield non-trivial dispersion values, there is often minimal improvement to the $\chi^2_r$ when compared to models without a depolarisation term (e.g. thin, DFR). This is especially true in the west lobe. We therefore argue that not all regions of NGC\,612  require a dispersion model (EFD, IFD). Indeed the strongest dispersion values are located in areas of the lobes that are best fit with a depolarisation term, in that the returned $\chi^2$ ~is lower. In the region surrounding the hot spot there are considerable amounts of Faraday dispersion, which increases in magnitude until it peaks at  the edge of our pixel sample.

\section{Discussion}
\label{n612:discuss}

It is not possible to definitively determine which polarisation mechanism is responsible for the observed Faraday rotation towards NGC\,612  using the $\chi_r^2$ statistic alone. The close $\chi^2_r$ values in Table~\ref{n612:chi2Table} and Figure~\ref{n612:chi2} show that all single-component models are capable of explaining the observed polarised signal at similar confidence levels. In the following subsections, we consider the physics of the observed Faraday depth signal in conjunction with the relative location of the thermal material along the line-of-sight. We apply any additional information available that may be able to help in distinguishing between models. The key difference in the interpretation of the polarisation models is the location of the Faraday rotating material. The Faraday simple models (thin, EFD) have the rotating material external to the synchrotron radio lobes whereas the Faraday rotation of the complex polarisation models (DFR, IFD) is taking place within the radio lobe. For the remainder of the discussion, we refer to the Faraday simple models as `Ext' and the complex models as `Int', symbolising where the Faraday rotation is taking place.

\subsection{Minimum Energy Estimates}
\label{n612:minEnergy}

We can estimate the minimum energy magnetic field needed to sustain the lobes of NGC\,612  using equation 4 from \citet{BeckKrause2005}. We break the lobes into two symmetric rectangular slabs each with a line-of-sight~pathlength of $140$\,kpc, and spectral index of $\alpha\,=\,-0.65$. At 2100\,MHz the eastern lobe has a surface brightness of $144$\,mJy/beam and an average polarised fraction of $23\%$. The western lobe has a surface brightness of $121$\,mJy\,beam$^{-1}$~and an average degree of polarisation of $16\%$. We assume that the polarised emission comes from a regular field with all possible inclinations and that the synchrotron plasma has a filling factor ($f$) of $0.1$ throughout the lobes. We have no information on the proton-to-electron ratio (${\bf K_{0}}$), therefore we assume unity. With these estimates in mind, we find minimum energy magnetic field strengths of $4.3$ and $4.2$\,$\mu$G for the eastern and western lobes respectively. These estimates in turn lead to energy densities of $2.0\times10^{-13}$ and $1.8\times\,10^{-13}$\,erg\,cm$^{-3}$ for the eastern and western lobes.  The field strength is not highly dependent on the inclination of the magnetic field under the assumption that the angle is averaged over the entire synchrotron volume of the lobes. On the other hand, we note that these estimates depend strongly on the filling factor ($f$) and the ${\bf K_{0}}$-value. 

Our average minimum energy estimate of $\overline{B}\,\sim\,4.2\,\mu$G is not within the range provided by \citet{Tashiro2000}, who use diffuse X-ray emission and find an implied field strength of  $B\,\simeq\,1.6\,\pm{1.3}\,\mu$G for both lobes. Exploring the possibility that the variation between values could be a result of our estimation of effective pathlength and ${\bf K_{0}}$, we evaluate the minimum-energy magnetic field for decreasing $f$ and increasing  ${\bf K_{0}}$. Each iteration results in a larger estimation of the implied magnetic field strength and we are unable to mediate the discrepancies between our estimate of $\overline{B}$ and that reported by \citet{Tashiro2000} with this method. We note that equipartition field strengths can be overestimates of the true magnetic field strength by up to a factor of 3 \citep{Croston2005}. If our measurement of the magnetic field is indeed an overestimate, a correction factor of $0.3$ reconciles the differences between our $\overline{B}$ and that of \citet{Tashiro2000}.

\subsection{External versus Internal Faraday Rotation}
\label{n612:RMloc}

\subsubsection{The Milky Way Foreground}
One possibility is that the observed Faraday rotation is dominated by a local Galactic component. At a Galactic latitude of $-77^{\circ}$, the Galactic contribution has been estimated to be on the order of a few rad\,m$^{-2}$ \citep{Oppermann2015}. The magnitude of this Galactic Faraday depth estimate is of the same order as the observed Faraday depth across the radio lobes ($\overline{\phi}_{ext}\,=\,3.3$\,rad\,m$^{-2}$; $\overline{\phi}_{int}\,=\,6.5$\,rad\,m$^{-2}$). Therefore, it is necessary to further investigate and characterise the Faraday contribution from the Milky Way. 

\citet{Stil2011} show that the Faraday depth of the Milky Way is coherent on angular scales of a few degrees. NGC\,612  has an angular scale of $\sim18'$ across the two radio lobes, suggesting that there would be minimal variation in the projected Faraday depth on the radio galaxy. However, Figure~\ref{n612:rmMaps} shows multiple regions on NGC\,612  where the intensity of the Faraday depth change over scales of a few arcminutes. 

Many of the areas where the Faraday depth is seen to change correspond to regions of interest that are local to NGC\,612. Figure~\ref{n612:rmMaps} shows that in the East lobe, the magnitude of the Faraday depth decreases as a function of  distance from the optical counterpart. In the West lobe, the Faraday depth is seen to increase in magnitude leading up to a ridge of depolarisation, first seen in Figure \ref{n612:polSpectra}. The most convincing trend in the Faraday depth is in the region of the hot spot, where the sign of the Faraday depth is observed to change. We will discuss the specific location of the hot spot in more detail in a subsequent section ($\S$\ref{n612:HS}), but the morphological correlation between $\phi$ and Stokes $I$ strongly suggest that the bulk of the observed Faraday rotation is taking place within or near the lobes of NGC\,612 . It follows that while the Galaxy is responsible for some amount of Faraday depth, it is unlikely that it is responsible for the bulk of the observed Faraday rotation.

\subsubsection{The ambient X-ray IGrM}

\citet{Guidetti2011, Guidetti2012} have argued that the Faraday rotation associated with radio lobes is due to the radio galaxy being  embedded in a halo of thermal material. In these instances, the galaxies investigated are members of galaxy clusters and the radio lobes were thought to be embedded in a halo of hot X-ray gas. The galaxy group environment often lacks a diffuse X-ray component; however, \citet{Tashiro2000} observe excess diffuse, soft X-ray emission extending $\sim200$\,kpc away from NGC\,612 , and argue the X-rays have been emitted via the IC process, signifying the presence of free electrons in the intragroup medium.

If we assume that the diffuse material in which NGC\,612  is embedded is threaded with a coherent magnetic field, then na{\"i}vely, any change in the distribution of the X-ray-emitting gas would correlate with a change in the observed Faraday rotation. This assumption of the foreground magnetic field also implies that any regions that are devoid of X-ray emitting gas will correspond with a Faraday depth signal that is consistent with zero. \citet{Tashiro2000} point out a clear anisotropy in the soft X-ray intensity with the majority of emission being associated with the eastern lobe. By contrast, the observed Faraday depth signal is largely isotropic across the two lobes. If the radio galaxy were embedded in a volume of hot gas, the apparent uniformity of the Faraday depth signal becomes difficult to explain given the lopsided nature of the X-ray emission. While this discrepancy does not serve as direct evidence against Faraday rotation due to a magnetised intragroup medium, it does raise some intriguing concerns as to what is responsible for the observed rotation. Additional high resolution X-ray imaging of the hot gas on the scale of the lobes is needed for further analysis of the potential for external Faraday rotation due to the immediate environment. 

\subsubsection{A swept up thin skin}
It is also possible that the expanding lobes of NGC 612 have swept up and compressed the surrounding intragroup medium, as has been argued to be the case by \citet{RudnickBlundell2003}, although limitations in their specific approach are detailed in \citet{Ensslin2003}. In this scenario, the observed Faraday depth signal is due to the synchrotron lobes being girt by a thin skin of thermal material. 

We explore this possibility by assuming such a boundary layer has a depth $dl\sim20$\,kpc; which is approximately one-tenth of the scale-height of the lobes. In $\S$\ref{n612:minEnergy}, we estimate the average total magnetic field in the lobes of NGC\,612  to be $\overline{B}\sim\,4.2\,\mu$G. As we have no knowledge of the relative field strengths of the line-of-sight component, we assume a geometrical upper limit of $B_{\parallel}\,\sim\,B/\sqrt{3}\,\sim\,2.4\,\mu$G. Using a mean Faraday depth of $\overline{\phi}_{\rm{EFD}}=\,+3.3$rad\,m$^{-2}$, we derive an implied electron density for a thin skin of $n_e\,\simeq\,8\times10^{-5}$ cm$^{-3}$ ($\phi/3.3$\,rad\,m$^{-2}$) ($B_\parallel/2.4\mu$G) ($dl/20$\,kpc). 

 \begin{figure}
\centering
\includegraphics[width = \linewidth]{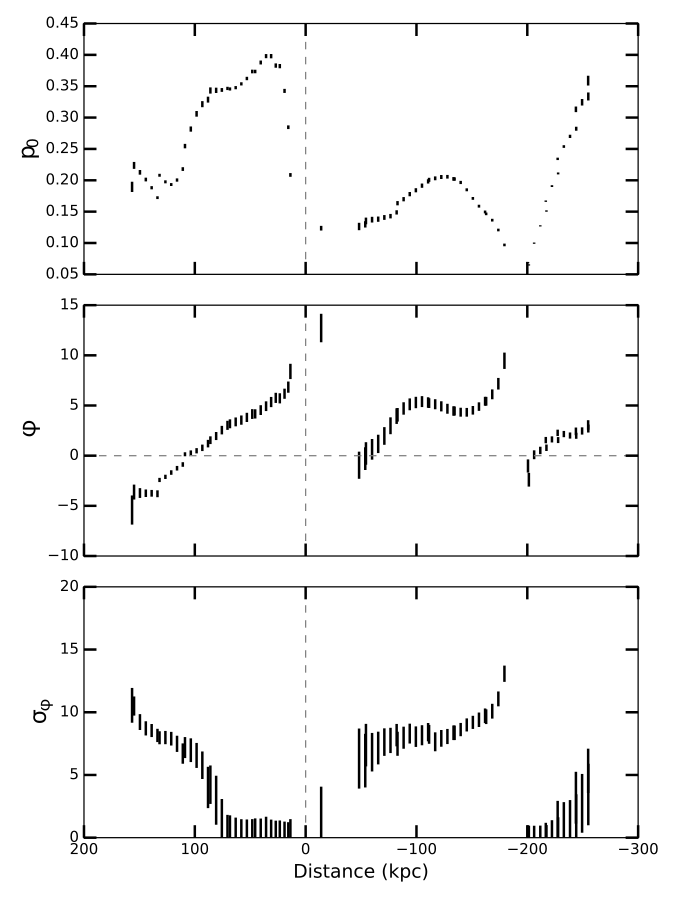}
	\caption[Profiles of various polarisation parameters along the jet of NGC\,612.]{Intrinsic polarisation, Faraday depth and Faraday dispersion as a function of position along the lobes of NGC\,612, as traced by our assumed jet path (e.g. Figure~\ref{n612:p0}. The length of each tick mark represents the corresponding uncertainty at each position. The $x$-axis is indicative of the distance from the optical galaxy, with negative values indicating positions west of the galaxy. The vertical dashed line therefore marks the position of the host galaxy. The horizontal dashed line in the middle plot indicates $\overline{\phi}\,=\,0$\,rad\,m$^{-2}$. }
	\label{n612:jetRM}	
\end{figure}

By comparison, \citet{OSullivan2013} find an implied density of $n_e\sim\,1.5\times10^{-3}$\,cm$^{-3}$ for the lobes of Centaurus A, assuming the same skin-depth ($dl\,\sim\,20\,$kpc). They argue that the compression of material in the intragroup medium alone is not able to account for such a high density. By contrast, our estimate of the electron density of a thin-skin Faraday screen is a factor of ten less. An intragroup medium equivalent to our required density is typical of many galaxy groups \citep{MulchaeyZabludoff1998, Sun2012} and accumulating a population of ionised material of this density from the surrounding medium seems plausible, given a moderate ionisation fraction as insinuated by the surrounding X-ray IGrM. Many theoretical models of the evolution of radio lobes assume that the surrounding medium has zero magnetisation (e.g. \citealt{Gourgouliatos2010}) and typical observed magnetisation levels in an intragroup medium have been quite low at a radius of hundreds of kiloparsec from a galaxy group centre. However, \citet{Tashiro2000} argue that the magnetic energy density of the diffuse, large-scale X-ray plasma is roughly equal to the electron energy density of the lobes of NGC\,612 . In order to accumulate a density of a thin skin of $n_e\,\sim\,8\times\,10^{-5}\,$cm$^{-3}$, little compression of X-ray plasma is required. 

We note that our assumed thin-skin pathlength is largely speculative and motivated by similar values used for other radio lobes and the derived $n_e$ used in the previous argument is inversely proportional to the pathlength of the thin skin. To explore how this may change our argument of the thin-skin approximation, changing the pathlength by a factor of $2$ (i.e. $10$ kpc $\lesssim dl \lesssim 40$ kpc) results in an implied electron density within the range of $2\times\,10^{-4}$ cm$^{-3}\lesssim n_e \lesssim 4\times10^{-5}$ cm$^{-3}$. This range of electron densities would still be physically achievable through the compression of the X-ray IGrM. 

Given the lopsided distribution of X-ray plasma, the apparent uniformity of the Faraday depth signal in NGC\,612  might be explained if the expanding lobes have swept up the ambient IGrM as they expand outward. Hydrodynamic simulations carried out by \citet{Bicknell1990} show that as lobes expand, it is possible for Kelvin-Helmholtz instabilities to form on the surface of the lobe due to feedback between the synchrotron material and the surrounding medium. In such a case, waves of material would appear on the surface of the radio galaxy, resulting in a unique Faraday depth pattern. \citet{Bicknell1990} also point out that a positive and negative variation in the sign of the Faraday depth should exist on a scale length equivalent to half of the wavelength of the eddy. \citet{Anderson2017} have also studied such Faraday rotation signals for the lobes of Fornax A. We adapt a similar approach in exploring the possibility that the eddies can be physical structures below.

Assuming that the jet is the dominant driving force for the expansion of the radio lobes, we choose to evaluate the polarisation properties along the path that the jet traces, as shown in Figure~\ref{n612:rmMaps}. The middle panel of Figure \ref{n612:jetRM} shows a rough sinusoid in the Faraday depth values as a function of distance from the optical counterpart. The half-wavelength of the surface wave appears to be $(\lambda/2)\sim\,1.1\,\rm{arcmin}\,\sim40$\,kpc. Although in all instances we do not see the Faraday depth change when the gradient of the Faraday depth changes direction, which is expected from Faraday eddies, we note that we have not corrected for the Faraday rotation contribution from the Milky Way, which may be shifting the observed Faraday depth to more positive values.

If we assume that the depth of the surface wave ($d_e$) is related to the wavelength of the eddy ($\lambda_e$), by a ratio of  $d_e/\lambda_e\sim\,0.3$ \citep{Bicknell1990} it is possible to estimate $n_e$ and $d_e$ using Equation 5.2 from \citet{Bicknell1990},
\begin{equation}
	\Delta\phi \sim 0.49\,\times\,n_eB_\parallel\lambda_e,
\end{equation}
\noindent where $\Delta\phi$ is the change in Faraday depth between peaks. From Figure~\ref{n612:jetRM}, we can see that the peak-to-peak difference in Faraday depth is $\Delta\phi\,\simeq\,6$\,rad\,m$^{-2}$.  Using our previously discussed estimate of $B_\parallel\simeq2.4\,\mu$G and an eddy wavelength of $\lambda_e\,\sim\,80$\,kpc (centre plot, Figure \ref{n612:jetRM}), we find that the necessary electron density in the surface wave would be $n_e\simeq\,6\,\times10^{-5}$\,cm$^{-3}$ with an implied surface wave depth of $d_e\,\simeq\,20$\,kpc. 

This is in excellent agreement with our thin skin estimate using the mean Faraday depth, although there are large uncertainties in our assumptions. This exercise argues that the observed Faraday signal associated with the bulk of the lobes of NGC\,612  can be explained by Faraday eddies formed via an interaction with the X-ray IGrM. This scenario also indicates a preference for the Faraday rotating material to be located external to the synchrotron-emitting plasma.

\subsection{Internal Faraday Rotation}
\label{n612:internalRM}

In contrast to the previous section, we now explore the possibility that the thermal- and synchrotron plasmas are located cospatially within the radio lobes. 

NGC\,612  contains a young stellar population with on-going large-scale star-formation throughout the disc of the galaxy \citep{Emonts2008}. It is possible that as the lobes of NGC\,612  expanded through the disk of the optical galaxy, it advected some material from the stellar disk \citep{Begelman1989, Churazov2001}. 

The advection process may be responsible for the observed Faraday rotation since any thermal material in the lobes would have been entrained from the galactic disk \citep{LaingBridle2002}. In this instance, the synchrotron emitting plasma is mixed with the thermal material from the galaxy and the bulk of the observed Faraday rotation takes place internally to the radio lobes. If the Faraday rotation associated with NGC\,612  is taking place inside the lobes of the radio galaxy, it would be one of only a few objects (e.g. \citealt{OSullivan2013}) that have been shown to have this distribution of magneto-ionic material. It should be noted that the work carried out by \citet{OSullivan2013} was over a more limited frequency range ($1288\,-\,1480\,$MHz) with more sparse  sampling across the imaged bandwidth. As polarisation work towards multiple radio sources has yet to be carried out over as large of a band as the work presented here, it is possible that the detection of internal Faraday rotation could become more frequent towards large radio lobes. 

Assuming that the lobes are threaded with thermal, magneto-ionic material, it is possible to approximate the amount of thermal material that needs to be diffused in the the lobes of NGC\,612  to produce the observed signal. Using the definition of Faraday depth from \citet{Burn1966},
 \begin{equation}
\phi(L)~=~0.812\int^0_L~n_{e}B_{\parallel}dl,
\label{eq:RM}
 \end{equation}
\noindent we can solve for the free electron density ($n_e$).  We take $L$ to be the effective distance through the magneto-ionic material in parsecs, that is the pathlength ($l$) times the filling factor of ionized gas along the total line-of-sight ($f$) \citep{Reynolds1991}. Using our previous assumed values for the pathlength ($l$), filling factor ($f$) and $B_\parallel$ ($\S$\ref{n612:minEnergy}), we calculate an implied density of $n_{e}\,\sim\,1.8\,\times\,10^{-5}$\,cm$^{-3}$. The radio lobes can be roughly resemble two cylinders, each with a radius of 70\,kpc and lengths of 160 and 220\,kpc, for the eastern and western lobes, respectively. This geometry implies a total volume of the radio emission of $V\,\sim\,2\times10^{71}$\,cm$^{3}$. Were this volume to be uniformly filled with the above calculated $n_e$, it follows that the implied thermal mass within the lobes would be $M_{th}\,\sim\,n_{e}m_{H}fV \sim\,2.5\,\times\,10^{8}\,M_{\odot}$ where $m_{H}$ is the mass of ionised hydrogen. We note that this estimate depends heavily on the numerous assumptions that we have made.

It is possible that thermal material was entrained from the galaxy as the jet pushed its way through the galactic disk \citep{LaingBridle2002}. If we assume an age for the radio lobes of ~0.1\,Gyr \citep{Blundell2000, Parma2002}, the lobes of NGC\,612  would need to entrain an average amount of $\sim2.5\,M_{\odot}$\,yr$^{-1}$. This estimate is a few orders of magnitude larger than the amount needed to decelerate relativistic jets \citep{Bicknell1994, LaingBridle2002} and it is unlikely that entrainment is the sole origin for the bulk of the thermal material. It is more likely that the bulk of the magneto-ionic plasma responsible for the observed Faraday rotation signal has been accumulated by a combination of the mechanisms discussed (e.g. thin skin, Faraday eddies and entrainment).

\subsection{The Hot Spot}
\label{n612:HS}

The area surrounding the hot spot in the eastern lobe offers an intriguing polarisation signal. In this region, dispersion levels are strongest (Figure~\ref{n612:depolMaps}) and the Faraday depth sign is opposite to that of the majority of the radio galaxy (Figure~\ref{n612:rmMaps}). At this particular location, NGC\,612  is interacting with its neighbouring galaxy NGC\,619 via a tenuous H\textsc{i} bridge \citep{Emonts2008}. 

Depolarisation can be caused by change in strength or direction of the coherent magnetic field. It is therefore possible that the observed depolarisation is due to an increase in the turbulence of the magnetic field as the jet of NGC\,612  is ploughing into the tidal bridge. However, this increase in random motion would not give rise to a sign change in the coherent magnetic field direction. Furthermore, Figure~\ref{n612:psiMaps} shows that the intrinsic polarisation angle near the hot spot does not have a significant change in orientation, insinuating that the corresponding sign-change in Faraday depth does not come about due to a characteristic change in the geometry of the synchrotron plasma. Explaining the change from positive to negative Faraday depth may require additional factors. 

One possibility is that there exists an intervening cloud of magnetised gas along the line of sight associated with the hot spot. \citet{Emonts2008} detect a faint bridge of H\textsc{i} material spanning the distance between NGC\,612  and neighbouring galaxy NGC\,619. Recent work by \citet{Banfield2017} use the contrast between the hot spot and the larger diffuse lobes as evidence of a strong interaction of the radio galaxy with its surrounding environment. If this bridge, which is believed to be tidal in origin, were to host a coherent line-of-sight~magnetic field oriented opposite that of NGC\,612 , the observed Faraday depth associated with this region would appear opposite to the bulk of NGC\,612  if the magnetic field in the H\textsc{i} cloud were sufficiently negative. 

There have been numerous detections of magnetised, tidally stripped material associated with continuum bridges (e.g. \citealt{Condon1993, Nikiel2013-SQ, Basu2017}) as well as neutral H\textsc{i} tidal features \citep{Hill2009, McClureGriffiths-LA-2010, Kaczmarek2017}. The existence of a coherent magnetic field in the H\textsc{i} cloud could provide structural support against the intruding AGN jet, resulting in a longer lifetime for the tidal remnant. Higher resolution observations of the polarised signal associated with the HS, in addition to information on the free-electron density in the H\textsc{i} cloud, are needed to explore this region further.

\section{Conclusions}
\label{n612:conclude}

We have presented a spectropolarimetric study of the radio galaxy NGC\,612  using broadband observations covering 1.3 - 3.0~GHz taken with the ATCA. It is immediately evident that the polarisation signal depends upon the position within the lobes. We have demonstrated that majority of the polarisation signal seen associated with the radio lobes can be explained through a single polarisation component, although the exact mechanism remains unclear because of similarity between model fits. Using $qu$-fitting, we were able to confidently recover the intrinsic polarisation properties associated with the lobes.

We have explored any environmental implications from the observed Faraday depth signal. While we cannot rule out the entrainment of thermal material from the galactic disk, we find evidence of Faraday eddies on the surface of the synchrotron lobes, as seen by a sinusoidal variation in Faraday depth as a function of distance along the lobe. We argue that these surface waves are formed via the expansion of the synchrotron lobes into the surrounding medium, forming a thin skin ($dl\sim\,20$\,kpc) of thermal material that is responsible for the bulk of the observed Faraday rotation.

We estimate a total minimum magnetic field strength of $B_{min}\sim4.2\,\mu$G in the lobes of NGC\,612 . If the thermal material is distributed as a thin skin, we calculate a free-electron density in the skin of order $n_e\,\sim\,10^{-4}$\,cm$^{-3}$, assuming equipartition. As NGC\,612  is embedded in a halo of hot, X-ray emitting plasma, we argue that achieving this density would require little compression of the ambient IGrM. 
 
 We observe intriguing Faraday signal at the location of the hot spot in the east lobe of NGC\,612 .  At this location, a H\textsc{i} cloud, arising from a previous interaction, has been observed. We hypothesise that this signal might be explained if the tidal material is threaded with a coherent magnetic field, oriented in the opposite direction to the bulk of the Faraday rotating material. Future high-resolution polarisation observations, in addition to pathlength estimates, are needed to confirm this hypothesis.
 
 In order to correctly account for any spectral dependencies that may be associated with the two jets or the compact hot spot in the lobes, follow-up, high-resolution radio polarisation observations are required. Furthermore, high-resolution follow-up observations have the potential to break the ambiguity between our modelling results, in that they may resolve smaller surface features, that are predicted by our assumed thin skin model, which have been smoothed by our synthesised beam. Such an observational approach would be advantageous for all future related studies. The early science stages of the Australian Square Kilometre Array Pathfinder will provide additional $\lambda^{2}$ coverage allowing for the resolution of all Faraday components along the line-of-sight, making it easier to deduce the true polarisation mechanisms responsible for the observed signal. Better electron density estimates of the intragroup medium and magnetic field  estimates will strengthen further analysis of the NGC\,612  system and should be possible with sensitive X-ray observations, such as those offered with the XMM-Newton telescope.

 \section*{Acknowledgements}
 The Australia Telescope Compact Array is part of the Australia Telescope National Facility which is funded by the Australian Government for operation as a National Facility managed by CSIRO. We thank the staff at the Australia Telescope Compact array for their assistance and support with this project. We also appreciate the invaluable insight offered to us by  R. W. Hunstead, R. A. Laing and L. Rudnick. B. M. G. and C. R. P acknowledge the support of the Australian Research Council through grant FL100100114. The Dunlap Institute is funded through an endowment established by the David Dunlap family and the University of Toronto. N. M. M.-G. acknowledges the support of the Australian Research Council through Future Fellowship FT150100024. XHS is supported by the National Natural Science Foundation of China under grant No. 11763008.

 \bibliographystyle{mn2e}
\bibliography{allBibs}

\onecolumn
\appendix
\section{Polarisation parameter and uncertainty maps}
\label{n612:appendix}

In $\S$\ref{n612:modResults}, we have argued that each of the single-component polarisation models discussed in this work offer similar goodness-of-fit to the observed polarisation spectra. Below, we present each of the parameter maps returned from our model-fitting routine. The agreement in parameter values between all models adds credence to the data validity. We point out that the major discrepancies between the parameter maps lie in the maps of Faraday depth, in that in the case of internal Faraday rotation, the total Faraday depth is equal to two times the observed Faraday rotation ($\phi\,=\frac{1}{2}$RM).

We also present the corresponding uncertainty maps represent the average of the 1$\sigma$ deviations of the walkers above and below the resultant best-fit ($\sigma_+$ and $\sigma_-$, respectively). 

\clearpage

\begin{figure}
	\subfigure[$p_0$]{\includegraphics[width= 0.42\linewidth]{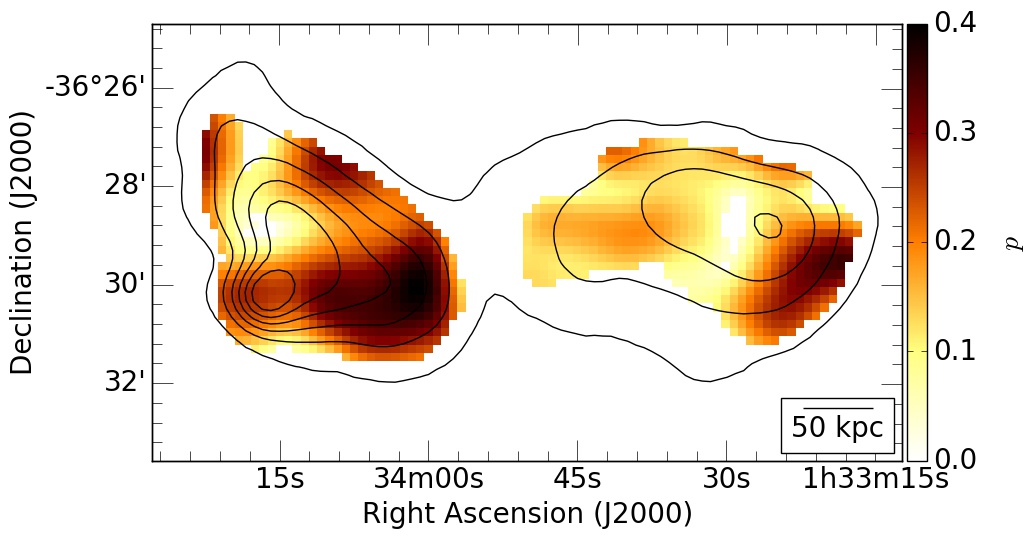}}
	\hfill
	\subfigure[$\psi_0$]{\includegraphics[width= 0.42\linewidth]{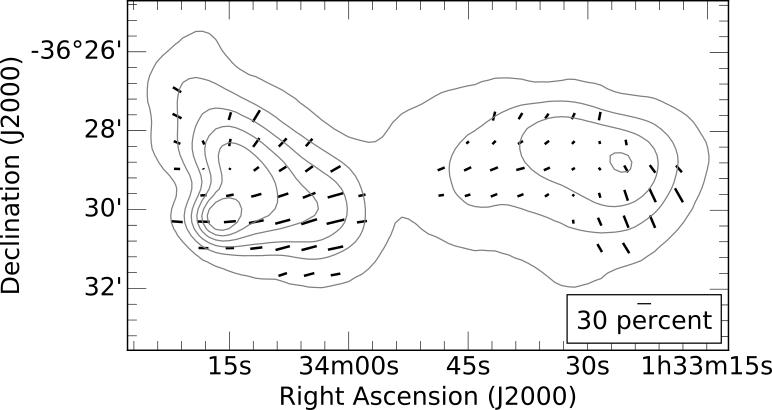}}
	\\
	\centering
	\subfigure[$\phi$]{\includegraphics[width= 0.42\linewidth]{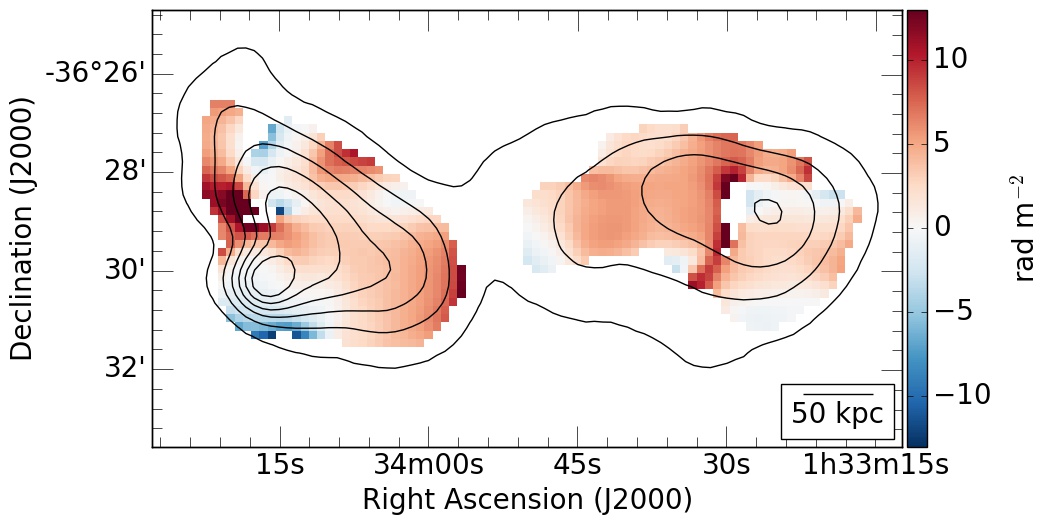}}
\caption[Parameter maps returned from fitting a Faraday thin model to NGC\,612]{Faraday thin parameter maps showing intrinsic polarisation (a), intrinsic polarisation angle (b) and Faraday depth (c). The length of each vector in Figure (b) represents the corresponding polarisation fraction at that location. Each subfigure has a scale bar in the lower righthand corner. Total intensity contours mark $25 - 400$\,mJy/beam every $75\,$mJy/beam. }
\label{n612:thin_parMaps}
\end{figure}

\begin{figure}
	\subfigure[$\sigma(p_0)$]{\includegraphics[width= 0.42\linewidth]{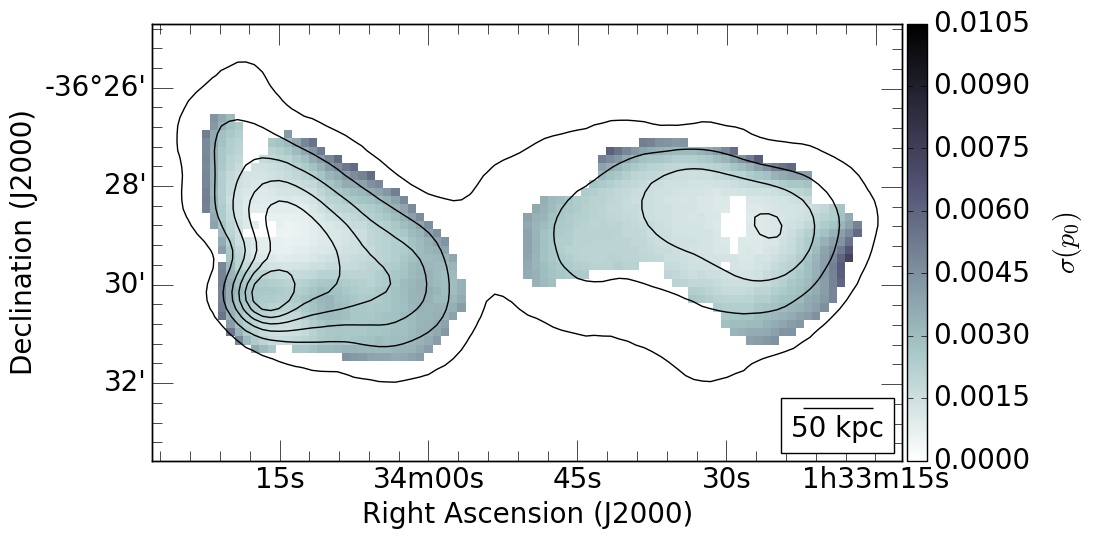}}
	\hfill
	\subfigure[$\sigma(\psi_0)$]{\includegraphics[width= 0.42\linewidth]{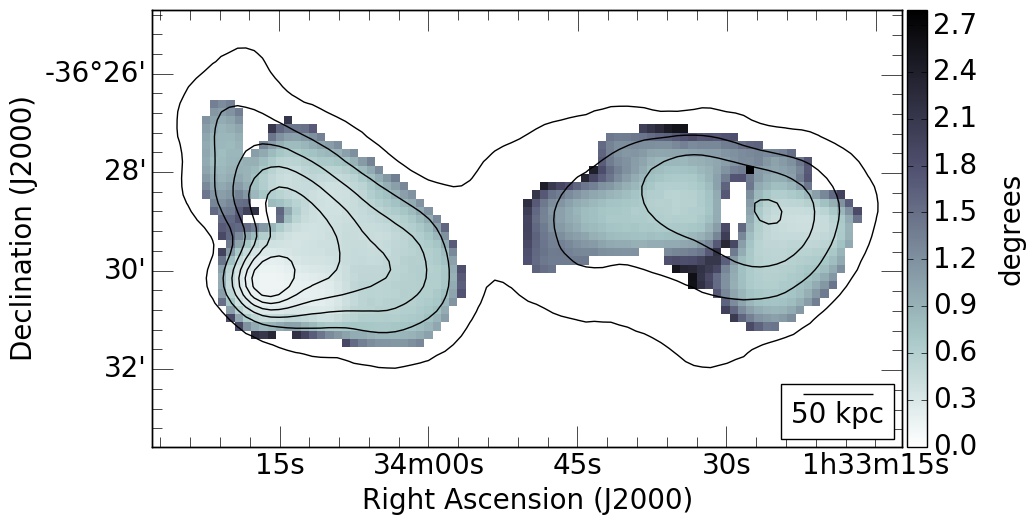}}
	\\
	\centering
	\subfigure[$\sigma(\phi)$]{\includegraphics[width= 0.42\linewidth]{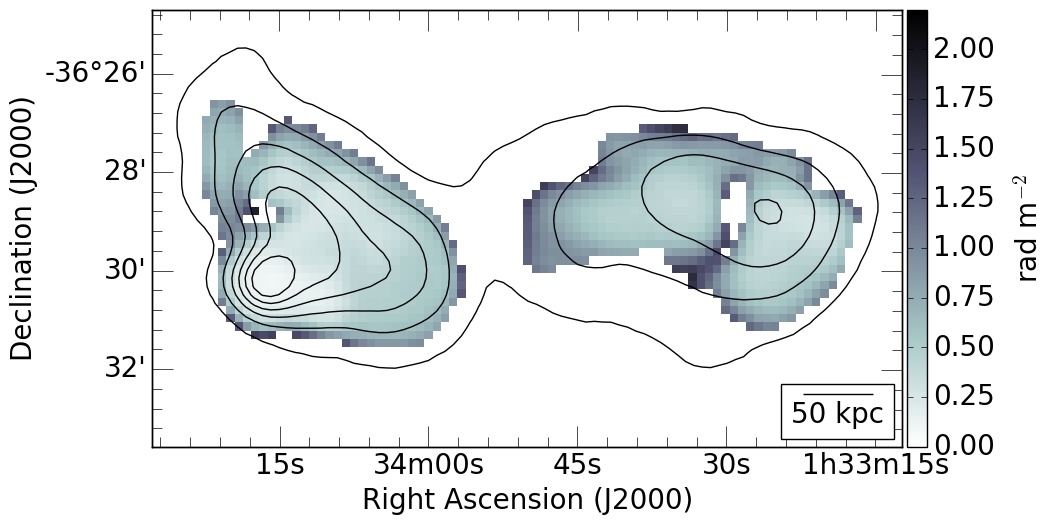}}
\caption[Parameter uncertainty maps returned from fitting a Faraday thin model to NGC\,612]{Uncertainty maps for a Faraday thin model. Maps shown correspond to uncertainty in intrinsic polarisation fraction (a), uncertainty in intrinsic polarisation angle, in degrees (b), and uncertainty in Faraday depth (c). Each subfigure has a scale bar in the lower righthand corner. Total intensity contours mark $25 - 400$\,mJy/beam every $75\,$mJy/beam. }
\label{n612:thin_dparMaps}
\end{figure}

\begin{figure}
	\subfigure[$p_0$]{\includegraphics[width= 0.42\linewidth]{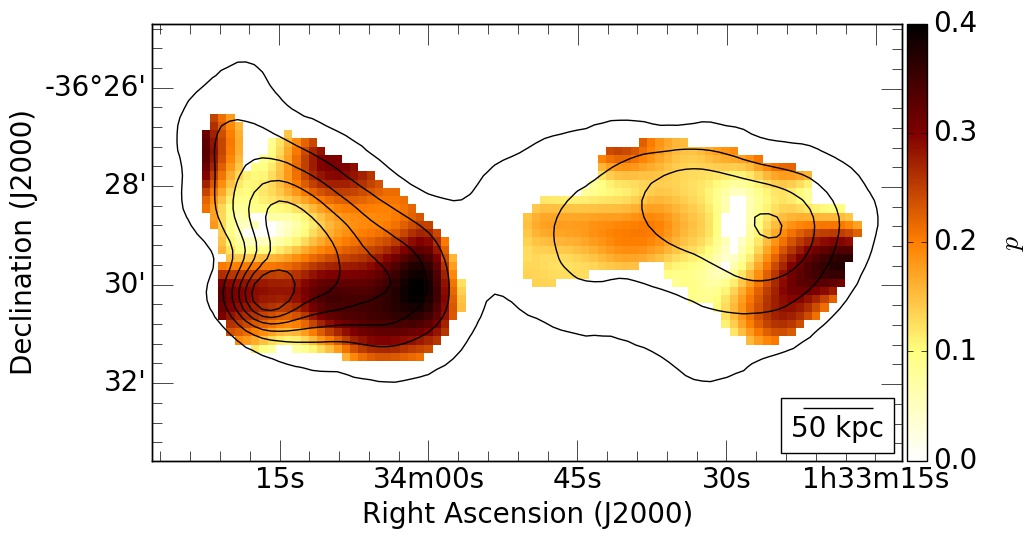}}
	\hfill
	\subfigure[$\psi_0$]{\includegraphics[width= 0.42\linewidth]{Figures/model2_Bvec.png}}
	\\
	\subfigure[$\phi$]{\includegraphics[width= 0.42\linewidth]{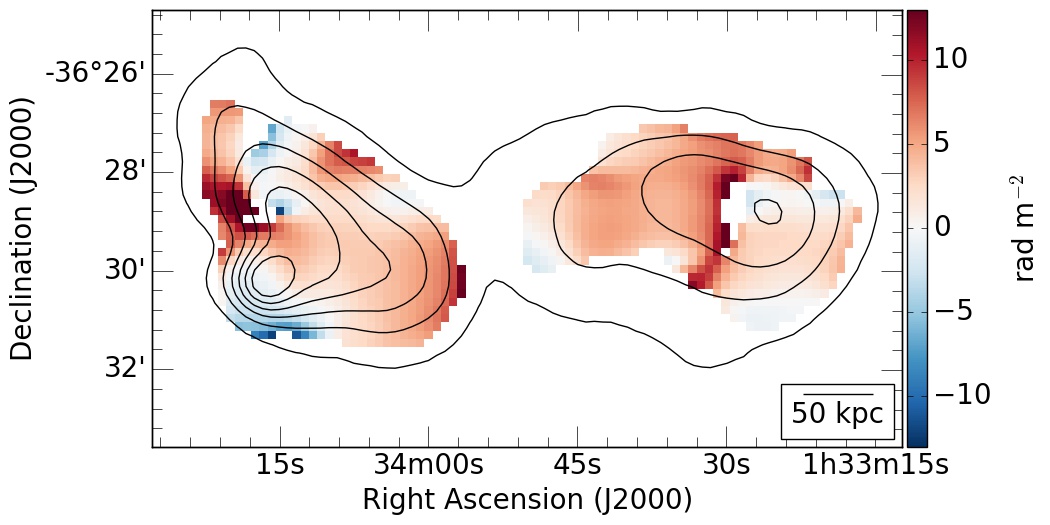}}
	\hfill
	\subfigure[$\sigma_{\phi}$]{\includegraphics[width= 0.42\linewidth]{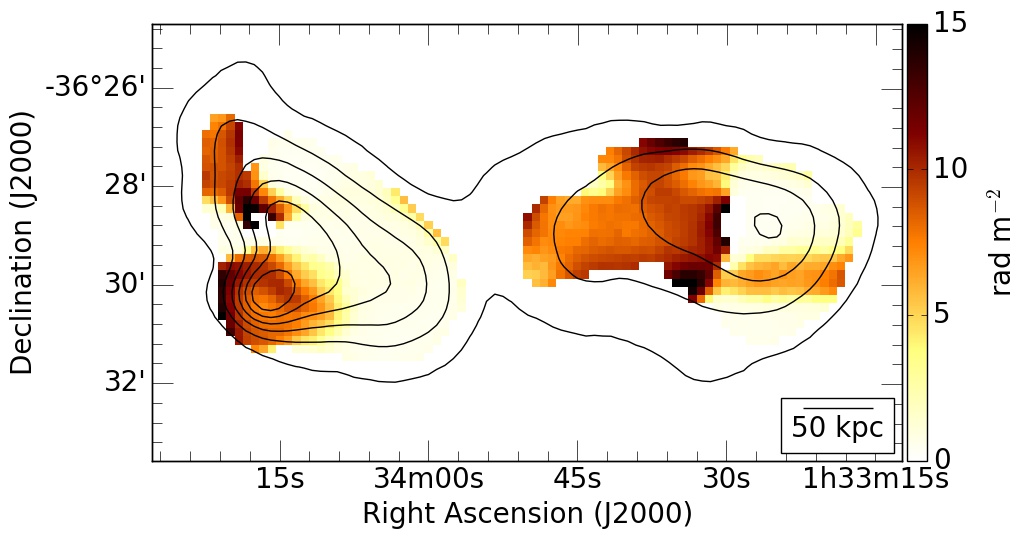}}\\
\caption[Parameter maps returned from fitting an external Faraday dispersion model to NGC\,612]{EFD parameter maps showing intrinsic polarisation (a), intrinsic polarisation angle (b), Faraday depth (c) and Faraday dispersion. The length of each vector in Figure (b) represents the corresponding polarisation fraction at that location. Each subfigure has a scale bar in the lower righthand corner. Total intensity contours mark $25 - 400$\,mJy/beam every $75\,$mJy/beam. }
\label{n612:EFD_parMaps}
\end{figure}

\begin{figure}
	\subfigure[$\sigma(p_0)$]{\includegraphics[width= 0.42\linewidth]{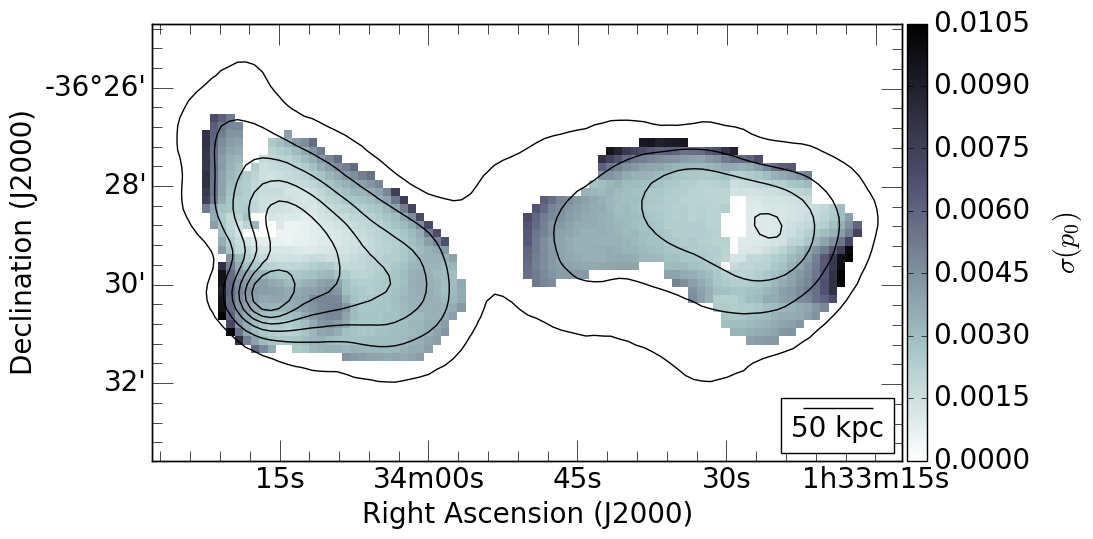}}
	\hfill
	\subfigure[$\sigma(\psi_0)$]{\includegraphics[width= 0.42\linewidth]{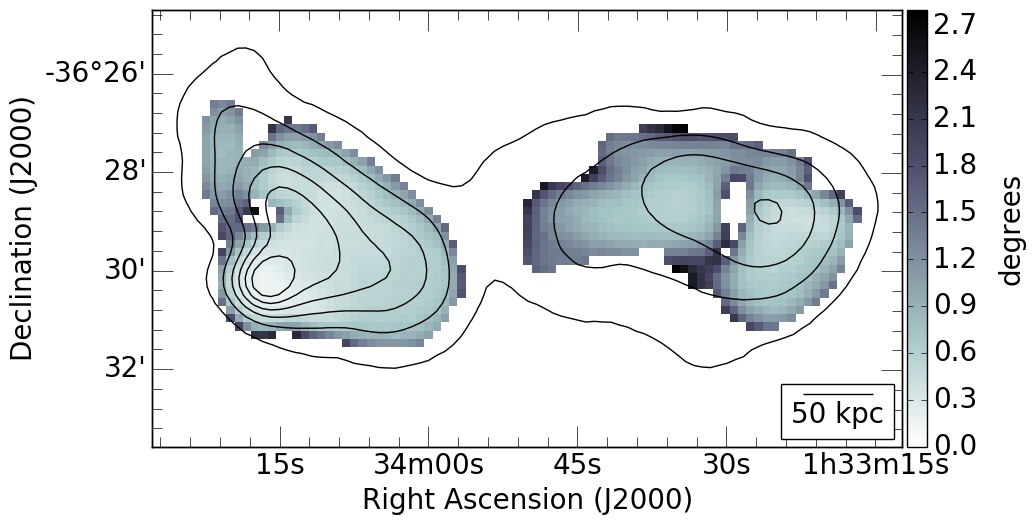}}
	\\
	\subfigure[$\sigma(\phi)$]{\includegraphics[width= 0.42\linewidth]{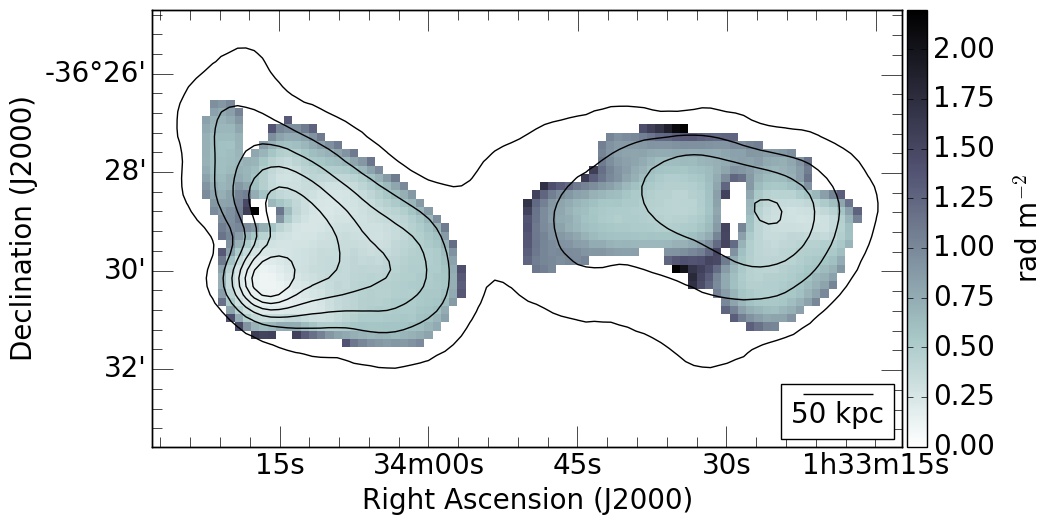}}
	\hfill
	\subfigure[$\sigma(\sigma_{\phi})$]{\includegraphics[width= 0.42\linewidth]{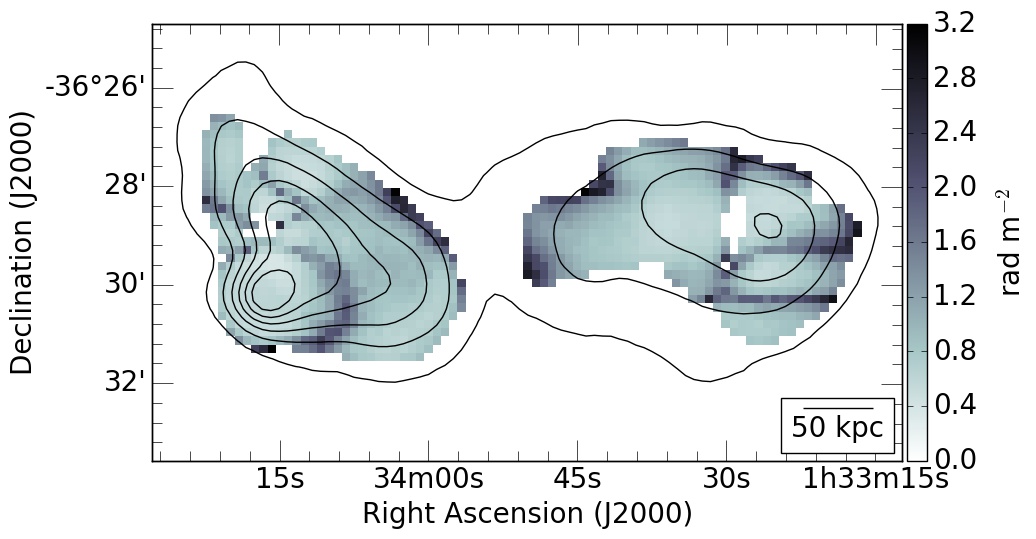}}\\
\caption[Parameter uncertainty maps returned from fitting an external Faraday dispersion model to NGC\,612]{Uncertainty maps pertaining to the EFD polarisation model. The figures shown represent uncertainty in intrinsic polarisation (a), uncertainty in the intrinsic polarisation angle (b), uncertainty in Faraday depth (c) and uncertainty in the Faraday dispersion (d). Each subfigure has a scale bar in the lower righthand corner. Total intensity contours mark $25 - 400$\,mJy/beam every $75\,$mJy/beam. }
\label{n612:EFD_dparMaps}
\end{figure}

\begin{figure}
	\subfigure[$p_0$]{\includegraphics[width= 0.42\linewidth]{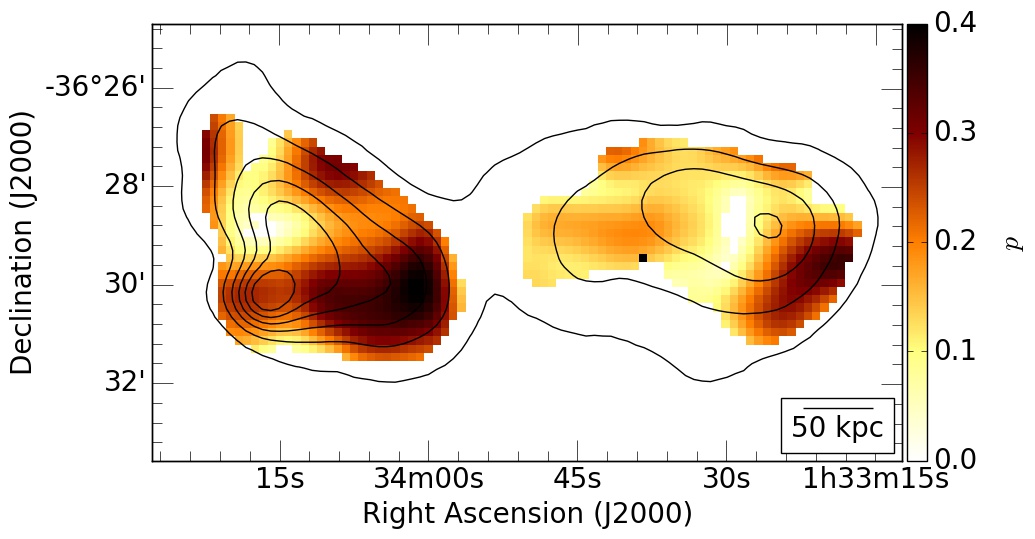}}
	\hfill
	\subfigure[$\psi_0$]{\includegraphics[width= 0.42\linewidth]{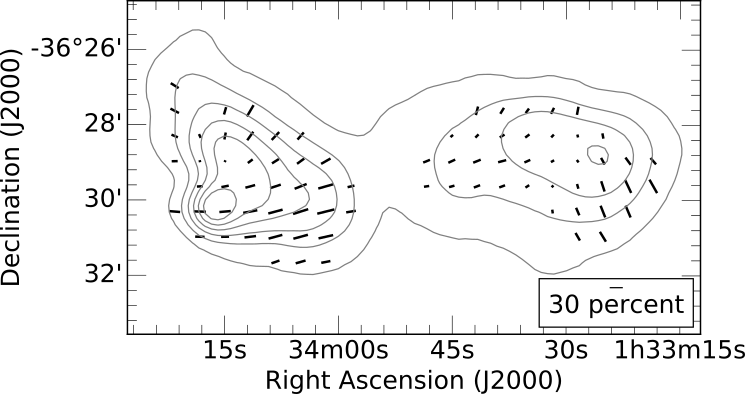}}
	\\
	\centering
	\subfigure[$\phi$]{\includegraphics[width= 0.42\linewidth]{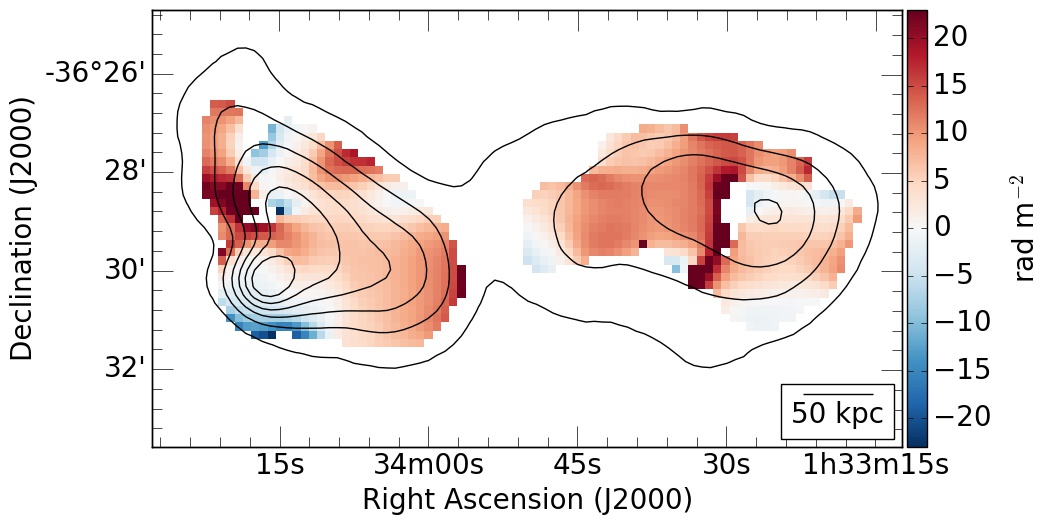}}
\caption[Parameter maps returned from fitting a differential Faraday rotation model to NGC\,612]{Parameter maps corresponding to the DFR polarisation model showing intrinsic polarisation (a), intrinsic polarisation angle (b) and Faraday depth (c). The length of each vector in Figure (b) represents the corresponding polarisation fraction at that location. Each subfigure has a scale bar in the lower righthand corner. Total intensity contours mark $25 - 400$\,mJy/beam every $75\,$mJy/beam. }
\label{n612:DFR_parMaps}
\end{figure}

\begin{figure}
	\subfigure[$\sigma(p_0)$]{\includegraphics[width= 0.42\linewidth]{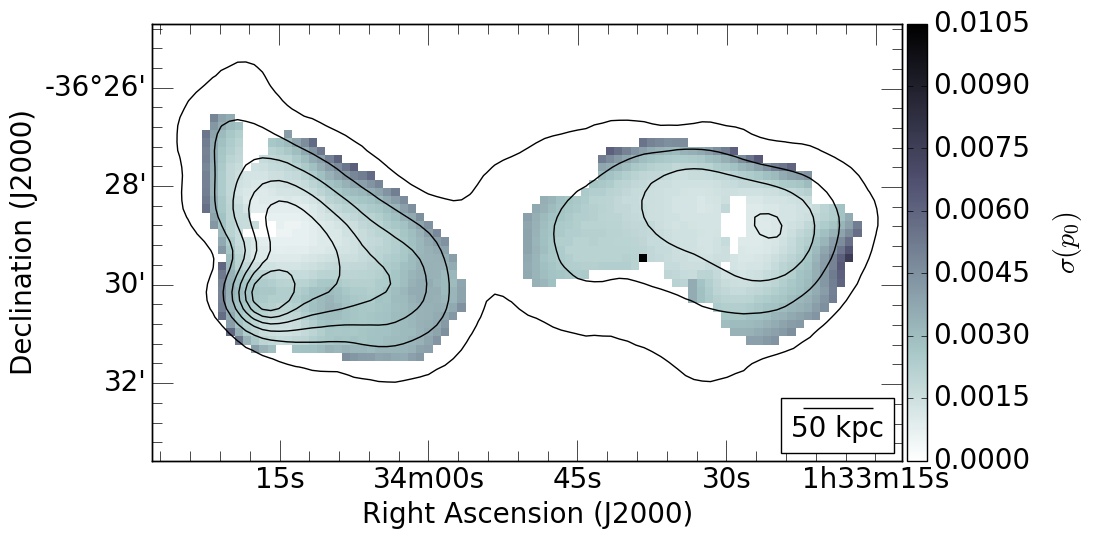}}
	\hfill
	\subfigure[$\sigma(\psi_0)$]{\includegraphics[width= 0.42\linewidth]{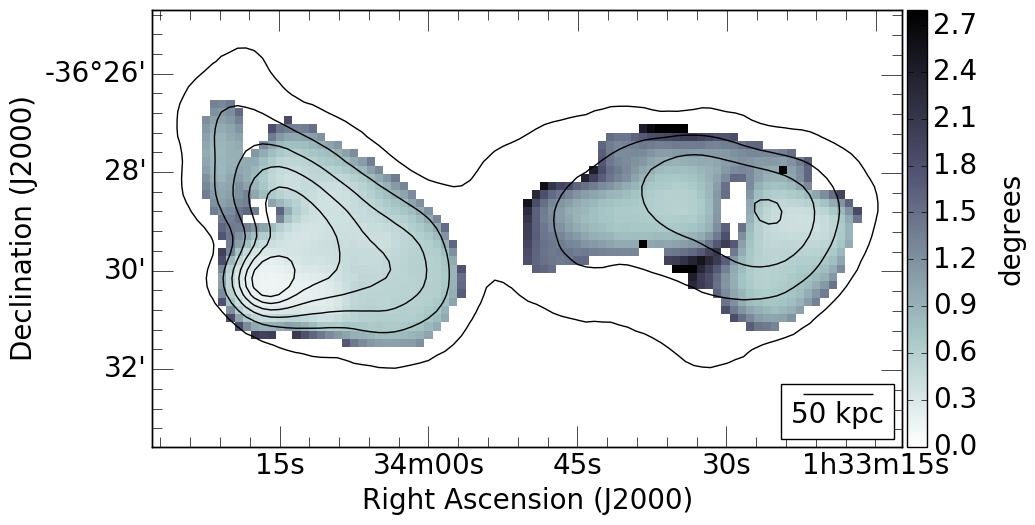}}
	\\
	\centering
	\subfigure[$\sigma(\phi)$]{\includegraphics[width= 0.42\linewidth]{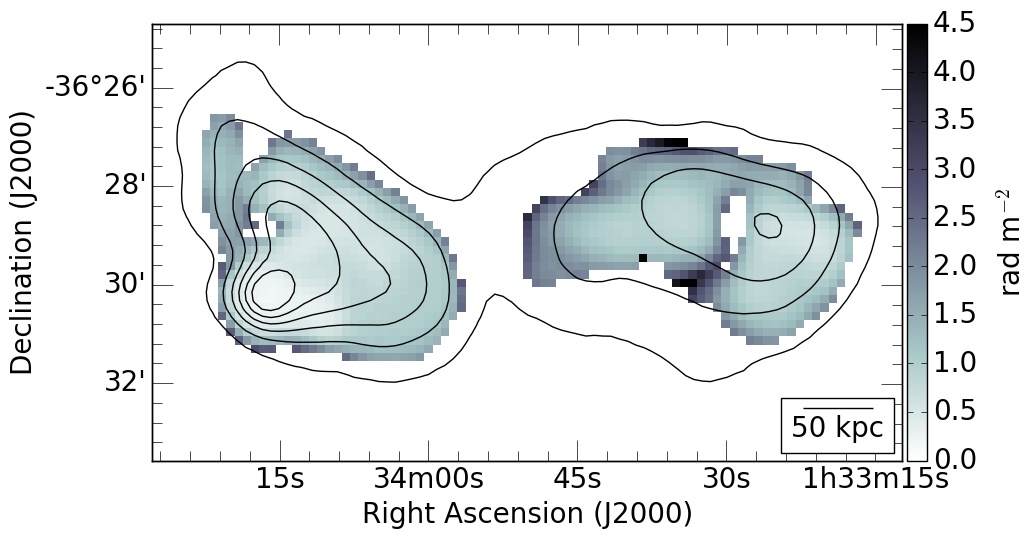}}
\caption[Parameter uncertainty maps returned from fitting a differential Faraday rotation model to NGC\,612]{Parameter uncertainty maps corresponding to the DFR polarisation model. Uncertainties shown are for intrinsic polarisation (a), intrinsic polarisation angle (b) and Faraday depth (c). Each subfigure has a scale bar in the lower righthand corner. Total intensity contours mark $25 - 400$\,mJy/beam every $75\,$mJy/beam. }
\label{n612:DFR_dparMaps}
\end{figure}

\begin{figure}
	\subfigure[$p_0$]{\includegraphics[width= 0.42\linewidth]{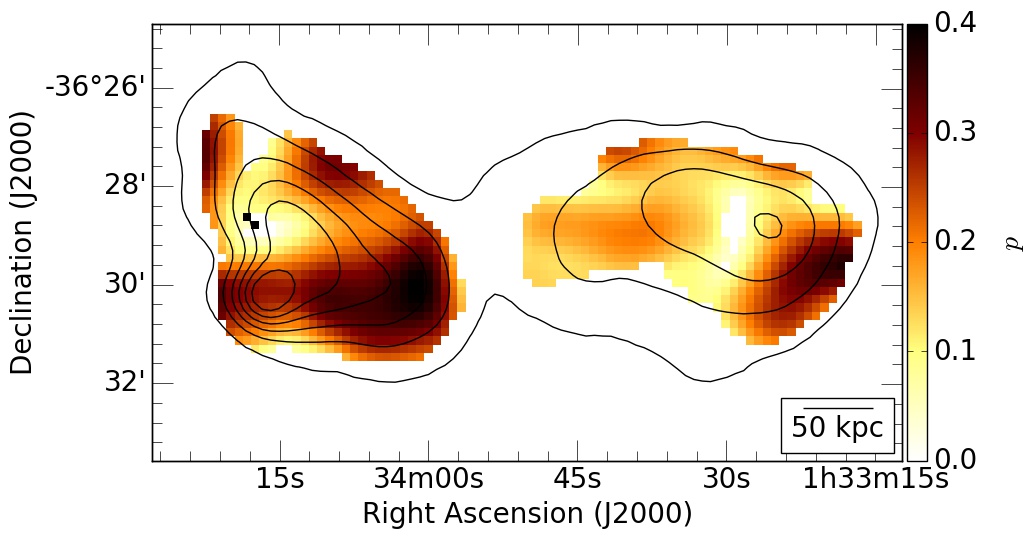}}
	\hfill
	\subfigure[$\psi_0$]{\includegraphics[width= 0.42\linewidth]{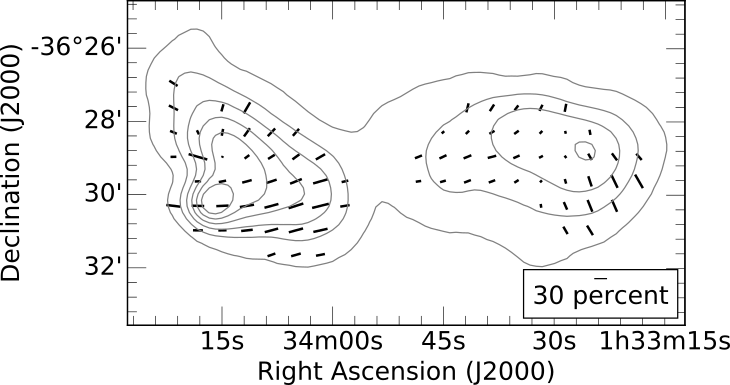}}
	\\
	\subfigure[$\phi$]{\includegraphics[width= 0.42\linewidth]{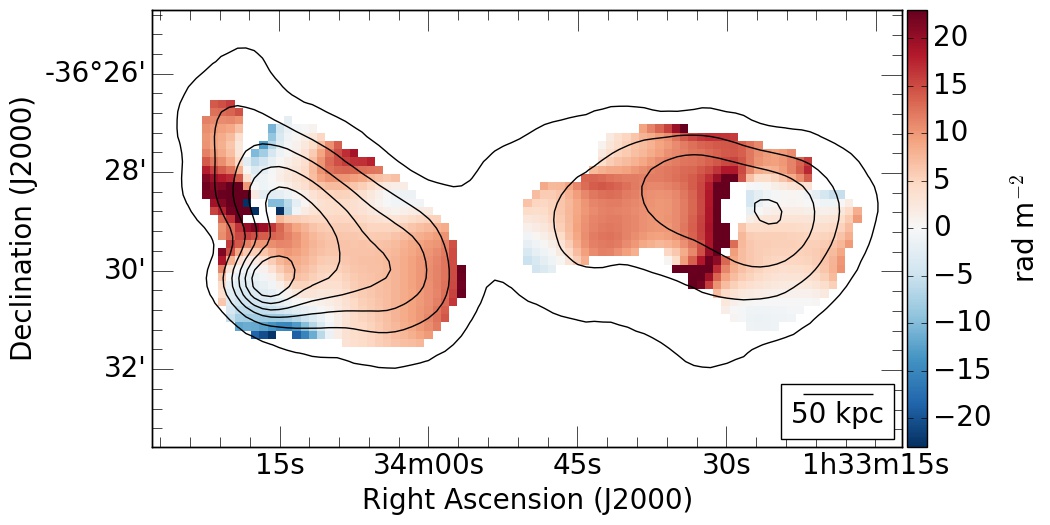}}
	\hfill
	\subfigure[$\zeta$]{\includegraphics[width= 0.42\linewidth]{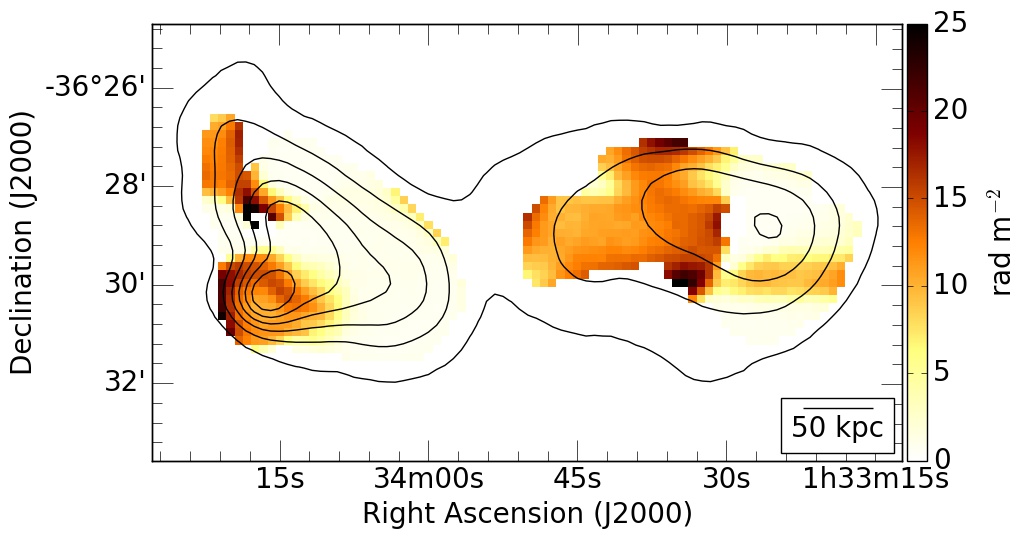}}\\
\caption[Parameter maps returned from fitting an internal Faraday dispersion model to NGC\,612]{Parameter maps corresponding to the IFD polarisation model showing intrinsic polarisation (a), intrinsic polarisation angle (b), Faraday depth (c) and internal Faraday dispersion (d). The length of each vector in Figure (b) represents the corresponding polarisation fraction at that location. Each subfigure has a scale bar in the lower righthand corner. Total intensity contours mark $25 - 400$\,mJy/beam every $75\,$mJy/beam. }
\label{n612:IFD_parMaps}
\end{figure}

\begin{figure}
	\subfigure[$\sigma(p_0)$]{\includegraphics[width= 0.42\linewidth]{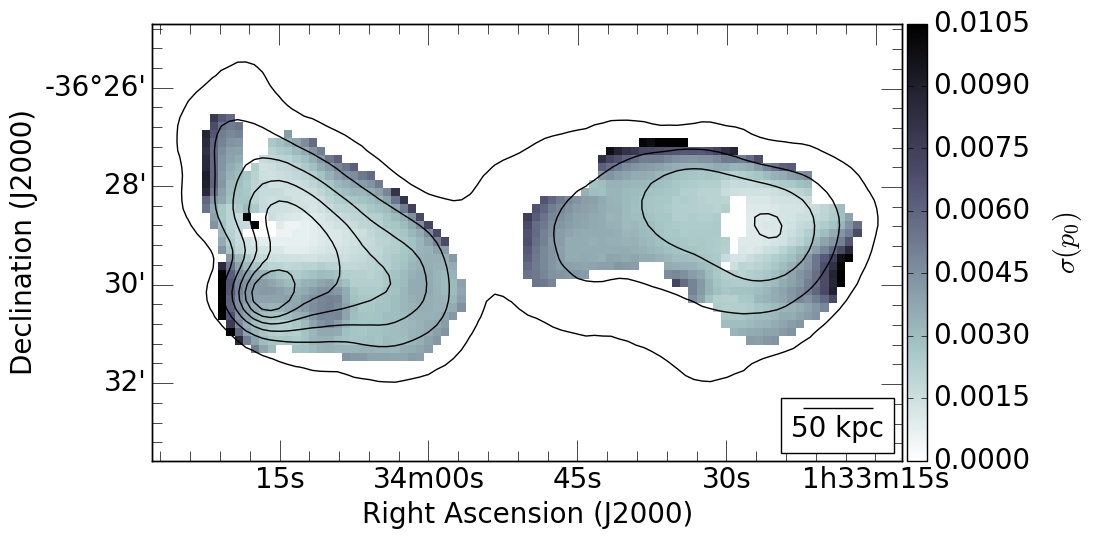}}
	\hfill
	\subfigure[$\sigma(\psi_0$]{\includegraphics[width= 0.42\linewidth]{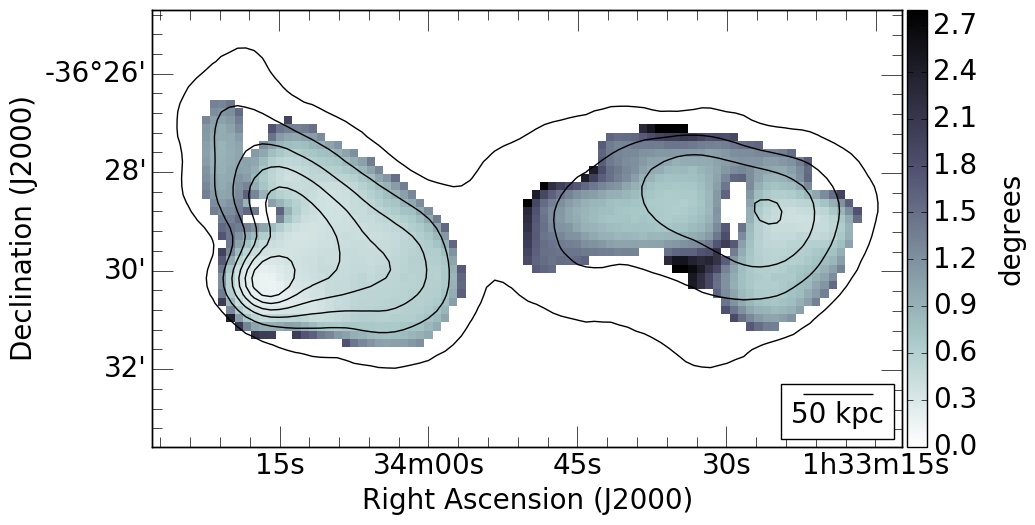}}
	\\
	\subfigure[$\sigma(\phi)$]{\includegraphics[width= 0.42\linewidth]{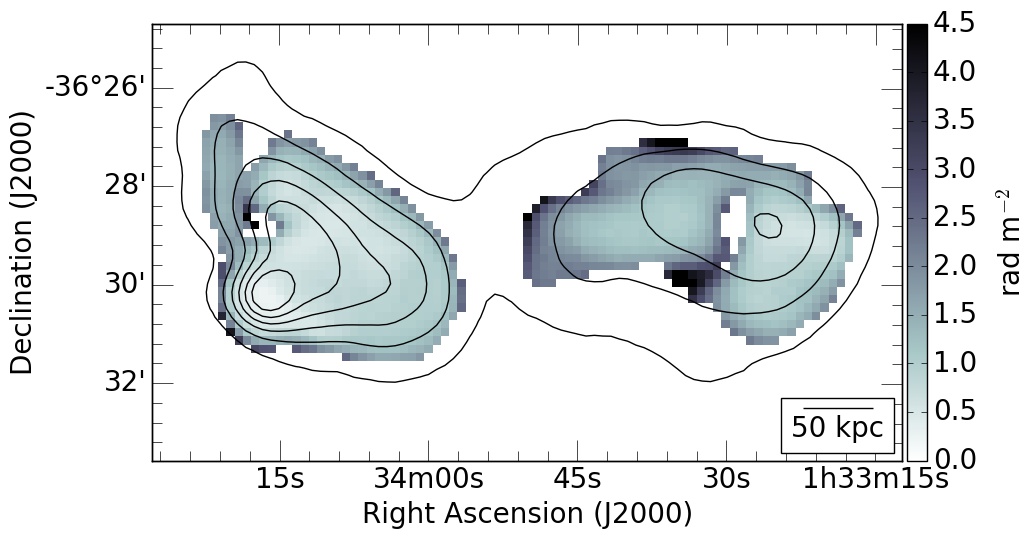}}
	\hfill
	\subfigure[$\sigma(\zeta)$]{\includegraphics[width= 0.42\linewidth]{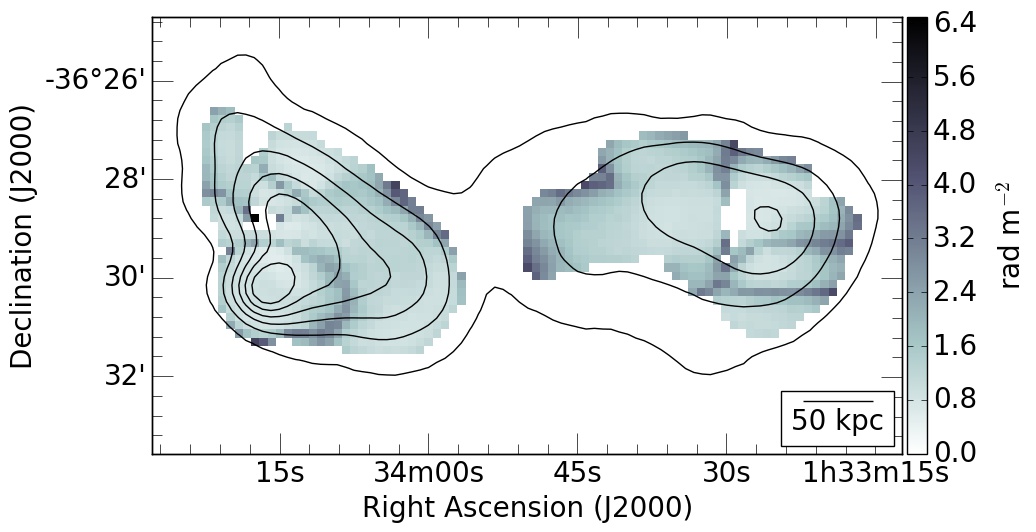}}\\
\caption[Parameter uncertainty maps returned from fitting an internal Faraday dispersion model to NGC\,612]{Parameter uncertainty maps corresponding to the IFD polarisation model. Uncertainties shown are for intrinsic polarisation (a), intrinsic polarisation angle (b), Faraday depth (c) and Faraday dispersion (d). Each subfigure has a scale bar in the lower righthand corner. Total intensity contours mark $25 - 400$\,mJy/beam every $75\,$mJy/beam.}
\label{n612:IFD_dparMaps}
\end{figure}

\label{lastpage}
\end{document}